\documentclass[12pt,twoside]{article}
\usepackage{verbatim}
\usepackage{epsfig}
\usepackage{epstopdf}
\input{epsf.sty}
\usepackage{epsf}
\usepackage{amssymb}
\usepackage{amsfonts}
\usepackage{amsmath}
\usepackage{pifont}
\usepackage[spanish,english]{babel}
\usepackage{graphicx}
\usepackage{subfigure}

\begin{document}

\title{Scalar coupling limits and diphoton Higgs decay from LHC in an $U(1)'$ model with scalar dark matter}
\author{R. Martinez$\thanks{%
e-mail: remartinezm@unal.edu.co}$, J. Nisperuza$\thanks{e-mail: jlnisperu@unal.edu.co}$, F. Ochoa$\thanks{%
e-mail: faochoap@unal.edu.co}$, J. P. Rubio$\thanks{e-mail: jprubioo@unal.edu.co}$, C.F. Sierra$\thanks{e-mail: cfsierraf@unal.edu.co}$ \and Departamento de F\'{\i}sica, Universidad
Nacional de Colombia, \\ Ciudad Universitaria, K. 45 No. 26-85, Bogot\'{a} D.C., Colombia}

\maketitle

\begin{abstract}
We explore constraints on the scalar coupling in a family nonuniversal $U(1)'$ extension of the standard model free from anomalies with a complex scalar dark matter particle. From unitarity and stability of the Higgs potential, we find the full set of bounds and order relations for the scalar coupling constants. Using recent data from the CERN-LHC collider, we study the signal strenght of the diphoton Higgs decay, which imposes very stringent bounds to the scalar couplings and other scalar parameters, including parameters associated to the dark matter. Taking into account these constraints, the observable relic density of the Universe, and the limits from LUX collaboration for direct detection, we obtain allowed masses for the dark matter particle as low as 55 GeV. By assuming that the lightest scalar boson of the model corresponds to the observed Higgs boson, we evaluate deviations from the SM of the trilineal Higgs self-coupling. The conditions from unitarity, stability and Higgs diphoton decay data allow trilineal deviations in the range $0 \leq \delta g \lesssim -72\%$.     

\end{abstract}

\section{Introduction}

After the observation of an 125 GeV scalar particle at CERN-LHC by the ATLAS and CMS collaborations  \cite{scalar_signal_ATLAS,scalar_signal_CMS}, the electroweak symmetry breaking mechanism has been experimentally stablished. Now, one of the highest priorities of the LHC experiments is to measure precisely the strenghts of the couplings of the Higgs boson to fermions and vector bosons \cite{couplingHiggsLHC}, which will allow to look for new states associated with the breaking symmetry mechanism in models beyond the standard model (SM) \cite{SM}. 
In particular, family nonuniversal $U(1)'$ symmetry models have many well-established motivations. For example, they provide hints for solving the SM flavor puzzle \cite{flavorphysics}, where even though all the fermions acquire masses at the same scale, $\upsilon =246$ GeV, experimentally they exhibit very different mass values. These models also imply a new $Z'$ neutral boson, which contains a large number of phenomenological consequences at low and high energies \cite{zprime-review}. 
In addition to the new neutral gauge boson $Z'$, an extended fermion spectrum is necessary in order to obtain an anomaly-free theory. Also, the new symmetry requires an extended scalar sector in order to {\it i.)} generate the breaking of the new Abelian symmetry and {\it ii.)} obtain heavy masses for the new $Z'$ gauge boson and the extra fermion content. Another consequence of an extended Higgs sector is that they may produce deviations of the Higgs self-coupling, which could provide an interesting test for the SM Higgs boson from future measurents at the LHC collider \cite{trilineal}.

On the other hand, the nonuniversal $U(1)'$ extension of the type introduced by authors in references \cite{U1} and \cite{U1DM}, it was proposed an extended scalar sector with two scalar doublets and two singlets with nontrivial $U(1)'$ charges, where the lightest scalar singlet is taken as candidate for scalar dark matter (DM) \cite{DM}. Some phenomenological consequences of this model have been studied in the above references, with special emphasis in the neutral gauge and Yukawa sectors. 

The main purpose of this paper is to determine some constraints on the parameters of the Higgs potential of the model, first by imposing theoretical bounds through unitarity and vacuum stability, and later by evaluating possible couplings of the observed Higgs boson to an extra scalar sector using experimental data at CERN-LHC. In particular, the signal strenght of Higgs boson decays to diphotons offers a clean signal to constraint new physics associated to extra scalar sectors, where one loop contributions from the charged Higgs bosons is taken into account. Also, since the diphoton signal strenght depends on the branching ratio with the total Higgs boson decay, it is possible to evaluate the effects of a light DM component as an invisible final state, where the scalar coupling to DM can be constrained. From these constraints and the limits from DM direct and indirect experiments, we obtain new limits on the mass values for light DM.   Finally, we obtain the Higgs self-coupling of the lightest Higgs boson in order to evaluate deviations from the SM prediction.

This paper is organized as follows. Section \ref{sec:model} is devoted to describe the spectrum and most important properties of the model. We also show the scalar couplings, including rotations into mass eigenvectors and trilineal interactions.  In section \ref{sec:unitarity}, we obtain constraints on the coupling constants of the Higgs potential from unitarity and stability conditions, where we use the whole space of parameters in order to obtain the most general bounds. In section \ref{sec:diphoton}, we obtain constraints in the space of parameters accessible to the observed decay of the Higgs boson to diphoton. For simplicity, in this section some parameters are taken to be equal. In section \ref{sec:DM} we discuss additional constraints from the DM direct and indirect experiments. Deviations to the SM Higgs self-coupling is evaluated in Section \ref{sec:trilineal}. Finally, we summarize our conclusions in section \ref{sec:conclusions}.    

\section{The Model\label{sec:model}}

\subsection{Particle content}

The particle content of the model \cite{U1} is composed of ordinary SM particles and new exotic non-SM particles, as shown in Tables \ref{tab:SM-espectro} and \ref{tab:exotic-espectro}, respectively, where column $G_{sm}$ indicates the transformation rules under the SM gauge group $(SU(3)_c,SU(2)_L,U(1)_Y)$, column $U(1)_X$ contains the values of the new quantum number $X$, and in the column labeled ``{\it Feature}'', we describe the type of field. Some properties of this spectrum are as follows:

\begin{enumerate}

\item The $U(1)_X$ symmetry is nonuniversal only in the left-handed SM quark sector: the quark family $i=3$ has $X_3=1/3$ while families $i=1,2$ have $X_{1,2}=0$. 



\item In order to ensure cancellation of the gauge chiral anomalies, the model includes in the quark sector three extra singlets $T$ and $J^{n}$, where $n=1,2$. They are quasichiral, i.e. chiral under $U(1)_X$ and vector-like under $G_{sm}$.

\item The most natural way to obtain massive neutrinos, according with the evidences of neutrino oscilations, is through a see-saw mechanism, which require the introduction of extra majorana neutrinos. Thus, to obtain a realistic model compatible with massive neutrinos, we include new neutral leptons $(\nu ^i_R)^c$ and $N ^i_R$.

\item An additional scalar doublet $\phi _2$ identical to $\phi _1$ under  $G_{sm}$ but with different $U(1)_X$ charges is included in order to avoid massless charged fermions, and 
where the individual vacuum expectation values (VEVs)  are related to the electroweak VEV through the relation $\upsilon = \sqrt{\upsilon  _1 ^2+\upsilon _2^2}$.

\item An extra scalar singlet $\chi _0$ with VEV $\upsilon _{\chi}$ is required to produce the symmetry breaking of the $U(1)_X$ symmetry. We assume that it happens at a large scale $\upsilon _{\chi} \gg \upsilon$.

\item Another scalar singlet $\sigma _0$ is introduced, which will be a DM candidate. Thus, this scalar must accomplish the following properties \cite{U1DM}: 

\begin{enumerate}

\item[(i)] Since $\sigma _0$ acquires nontrivial charge $U(1)_X$, it must be complex in order to obtain massive particles necessary for DM.

\item[(ii)] Terms involving odd powers of $\sigma _0$ induce decay of the DM, which spoils the prediction of the model for the DM relic density. Thus, we demand the following global symmetry

\begin{eqnarray}
\sigma _0 \rightarrow e^{i\theta }\sigma _0.
\label{global-symm}
\end{eqnarray}

\item[(iii)] In order to avoid the above symmetry to break spontaneously or new sources of decay, $\sigma _0$ must not generate VEV during the evolution of the Universe. Thus, we demand $\upsilon _{\sigma} = 0$.

\end{enumerate}

\item Finally, an extra neutral gauge boson $Z'_{\mu}$ is required to obtain a local $U(1)_X$ symmetry.

\end{enumerate}


With the above conditions, we construct the Higgs potential.

\subsection{Higgs potential}

The most general, renormalizable, $G_{sm} \times U(1)_X$ invariant potential and consistent with the global symmetry  (\ref{global-symm}) is

\begin{eqnarray}
V&=&\mu _1^2 \left| \phi _1 \right|^2 +\mu _2^2 \left| \phi _2 \right|^2 + \mu _3^2 \left| \chi _0\right|^2 +\mu _4^2 \left| \sigma _0\right|^2  \nonumber \\
&+&f_2\left(\phi _2^{\dagger} \phi _1 \chi _0+h.c.\right) \nonumber \\
&+& \lambda _1 \left| \phi _1 \right|^4+\lambda _2 \left| \phi _2 \right|^4+\lambda _3 \left| \chi _0\right|^4+\lambda _4 \left| \sigma _0\right|^4 \nonumber \\
&+& \left| \phi _1 \right|^2\left[ \lambda _6 \left| \chi _0\right| ^2+ \lambda '_6 \left| \sigma _0\right|^2 \right] \nonumber \\
&+& \left| \phi _2 \right|^2\left[ \lambda _7 \left| \chi _0\right| ^2+ \lambda '_7 \left| \sigma _0\right|^2 \right] \nonumber \\
&+&\lambda _5 \left| \phi _1 \right|^2\left| \phi _2 \right|^2+\lambda '_5 \left| \phi _1^{\dagger} \phi _2 \right|^2 +\lambda _8  \left| \chi _0\right|^2\left| \sigma _0\right|^2.
\label{higgs-pot-1}
\end{eqnarray}

As shown in \cite{U1DM}, the above potential lead us to the following mass eigenvectors:

\begin{eqnarray}
\begin{pmatrix}
G^{\pm}\\
H^{\pm}
\end{pmatrix}&=&R_{\beta}\begin{pmatrix}
\omega _1^{\pm}\\
\omega _2^{\pm}
\end{pmatrix}, \ \ \
\begin{pmatrix}
G_{0}\\
A_{0}
\end{pmatrix}=R_{\beta}\begin{pmatrix}
z_1\\
z_2
\end{pmatrix}, \nonumber \\
\begin{pmatrix}
h_{0}\\
H_{0}
\end{pmatrix}&=&R_{\alpha}\begin{pmatrix}
h_1\\
h_2
\end{pmatrix}, \ \ \
\begin{pmatrix}
H_{\chi }\\
G_{\chi }
\end{pmatrix}\sim I\begin{pmatrix}
h_3\\
z_3
\end{pmatrix},
\label{scalar-eigenvectors}
\end{eqnarray}
where $I$ is the identity, and the rotation matrices are defined according to

\begin{eqnarray}
R_{\beta, \alpha}&=&\begin{pmatrix}
C_{\beta, \alpha} & S_{\beta, \alpha} \\
-S_{\beta, \alpha} & C_{\beta, \alpha}
\end{pmatrix},
\end{eqnarray}
The rotation angles $\beta $ and $\alpha $ are:

\begin{eqnarray}
\tan \beta&=&T_{\beta}=\frac{\upsilon _2}{\upsilon _1},
\label{parameters}
\end{eqnarray}

\begin{eqnarray}
\sin{2\alpha }&\approx & \sin{2\beta}\left[1-\frac{\sqrt{2}C_{2\beta}S_{2\beta}\upsilon ^2}{f_2\upsilon _{\chi }}\left(\lambda _{1}C_{\beta }^2-\frac{\lambda _5+\lambda '_5}{2}C_{2\beta }-\lambda _{2}S_{\beta }^2\right) \right],
\label{alpha-angle}
\end{eqnarray}
while the eigenvalues for the dominant contributions are:

\begin{eqnarray}
M_{H^{\pm}}^2&\approx&M_{H_{0}}^2\approx M_{A_{0}}^2\approx-\frac{\sqrt{2}f_2\upsilon _{\chi}}{S_{2\beta }}, \nonumber \\
M_{H_{\chi }}^2 &\approx &2\lambda _{3} \upsilon _{\chi} ^2, \nonumber \\
M_{h_0}^2 &\approx& 2\upsilon ^2\left[\lambda _{1}C_{\beta }^{4}+\left(\lambda _{5}+\lambda _5'\right)C_{\beta }^{2}S_{\beta }^{2}+\lambda _{2}S_{\beta }^4\right].
\label{scalar-mass}
\end{eqnarray}


On the other hand, by assuming that the lightest scalar field $h_0$ corresponds to the observed Higgs boson, we are interested in the following trilineal couplings:

\begin{eqnarray}
V_{h_{0}}=g_{H^{\pm}}H^{+}H^{-}h_0+g_{\sigma }\sigma _0\sigma _0^{*}h_0+\frac{g_{3h}}{3!}h_{0}^{3},
\label{trilineal}
\end{eqnarray}
where the couplings are defined as

\begin{eqnarray}
g_{H^{\pm}}&=&\upsilon C_{\beta }\left(\lambda _5C_{\beta }^2C_{\alpha } +2\lambda _1S_{\beta }^2C_{\alpha }-\lambda _5'S_{\beta }C_{\beta }S_{\alpha }\right) \nonumber \\
&+&\upsilon S_{\beta }\left(\lambda _5S_{\beta }^2S_{\alpha } +2\lambda _2C_{\beta }^2S_{\alpha }-\lambda _5'S_{\beta }C_{\beta }C_{\alpha }\right), \nonumber \\ \nonumber \\
g_{\sigma }&=&\upsilon  \left(\lambda '_6C_{\alpha }C_{\beta }+\lambda '_7S_{\alpha }S_{\beta }\right), \nonumber \\ \nonumber \\
g_{3h}&=&6\upsilon \left[\lambda _1C_{\beta }C_{\alpha }^{3}+\lambda _2S_{\beta }S_{\alpha }^{3}+\frac{\lambda _5+\lambda '_{5}}{2}\left(C_{\beta }S_{\alpha }+S_{\beta }C_{\alpha }\right)C_{\alpha }S_{\alpha }\right]
\label{trilineal-coup}
\end{eqnarray}

\section{Theoretical constraints\label{sec:unitarity}}

First, we consider the theoretical constraints of the Higgs potential from unitarity and vacuum stability.

\subsection{Unitarity}

In order to calculate the tree unitarity bounds of the model, we use the LQT method \cite{key-1} developed by Lee,
Quigg and Thacker \cite{key-2}. It is based in the unitarity condition
of the $S$-matrix at tree level (through the optical theorem) and
the change  of the longitudinal components of
the massive vector boson fields by the respective Goldstone bosons in the limit at high energies
according to the equivalence theorem. This method has been used in the analysis
of two Higgs doublet models (THDM) in previous works \cite{key-3,key-4} and recently in an extended THDM
with an additional scalar singlet \cite{key-5}.

At high energies, the dominant contribution to the two-body scattering
processes comes from the quartic terms of the potential. Thus,
the unitarity bound for the $s$-wave amplitude of the $\mathcal{M}$-matrix
in the partial wave decomposition

\begin{equation}
\left|a_{0}\right|\leq\frac{1}{2},
\end{equation}
is reduced to the condition

\begin{equation}
\left|Q\right|\leq8\pi ,
\label{eq:bound}
\end{equation}
with $Q$ all the quartic couplings in the scalar sector. In order
to apply this condition, it is convenient to calculate the eigenvalues
of the $\mathcal{M}$-quartic matrix $Q$ in two particle processes. In our case, the quartic terms of the Higgs potential in Eq. (\ref{higgs-pot-1}) are:

\begin{align}
V_{4} & =\lambda_{1}\left|\phi_{1}\right|{}^{4}+\lambda_{2}\left|\phi_{2}\right|{}^{4}+\lambda_{3}\left|\chi_{0}\right|{}^{4}+\lambda_{4}\left|\sigma_{0}\right|{}^{4} \nonumber \\
& +\lambda_{5}\left|\phi_{1}\right|{}^{2}\left|\phi_{2}\right|{}^{2}+\lambda_{5}'\left|\phi_{1}^{\dagger}\phi_{2}\right|^{2}+\lambda_{8}\left|\chi_{0}\right|{}^{2}\left|\sigma_{0}\right|{}^{2} \nonumber \\
 & +\left|\phi_{1}\right|{}^{2}\left[\lambda_{6}\left|\chi_{0}\right|{}^{2}+\lambda_{6}'\left|\sigma_{0}\right|{}^{2}\right]\nonumber \\
 & +\left|\phi_{2}\right|{}^{2}\left[\lambda_{7}\left|\chi_{0}\right|{}^{2}+\lambda_{7}'\left|\sigma_{0}\right|{}^{2}\right],
\label{eq:quartic potential}
\end{align}
with the scalar field representations from tables \ref{tab:SM-espectro} and \ref{tab:exotic-espectro}.
%
%
In this way according with the LQT method, the $Q$-matrix can be
expressed as an $18\times18$ matrix with three independent block diagonal matrices
$\mathcal{M}_{1}(6\times6)$, $\mathcal{M}_{2}(9\times9)$ and $\mathcal{M}_{3}(3\times3)$
which do not couple with each other due to charge conservation and
CP-invariance \cite{key-3}. First, in the basis $(\omega_{1}^{+}\omega_{2}^{-},\omega_{2}^{+}\omega_{1}^{-},h_{1}z_{2},h_{2}z_{1},z_{1}z_{2},h_{1}h_{2})$
the symmetric submatrix $\mathcal{M}_{1}$ is given by:

\begin{equation}
\mathcal{M}_{1}=\left(\begin{array}{cccccc}
0 & \lambda_{5}+\lambda_{5}' & i\lambda_{5}'/2 & -i\lambda_{5}'/2 & \lambda_{5}'/2 & \lambda_{5}'/2\\
* & 0 & -i\lambda_{5}'/2 & i\lambda_{5}'/2 & \lambda_{5}'/2 & \lambda_{5}'/2\\
* & * & \lambda_{5}+\lambda_{5}' & 0 & 0 & 0\\
* & * & * & \lambda_{5}+\lambda_{5}' & 0 & 0\\
* & * & * & * & \lambda_{5}+\lambda_{5}' & 0\\
* & * & * & * & * & \lambda_{5}+\lambda_{5}'
\end{array}\right)
\end{equation}
with eigenvalues

\begin{align}
e_{1} & =\lambda_{5},\nonumber \\
e_{2} & =\lambda_{5}+2\lambda_{5}',\nonumber \\
f_{\pm} & =\pm\sqrt{\lambda_{5}(\lambda_{5}+2\lambda_{5}')},\nonumber \\
f_{1} & =f_{2}=\lambda_{5}+\lambda_{5}'.
\end{align}

The next basis of scattering processes corresponds to $(\omega_{1}^{+}\omega_{1}^{-},\omega_{2}^{+}\omega_{2}^{-},\frac{z_{1}z_{1}}{\sqrt{2}},\frac{z_{2}z_{2}}{\sqrt{2}},\frac{h_{1}h_{1}}{\sqrt{2}},\frac{h_{2}h_{2}}{\sqrt{2}},$\newline
$\frac{z_{3}z_{3}}{\sqrt{2}},\frac{h_{3}h_{3}}{\sqrt{2}},\sigma_{0}^{*}\sigma_{0})$
where the $\sqrt{2}$ factor accounts for identical particles, where:

\begin{eqnarray}
&&\mathcal{M}_{2}= \nonumber \\
&&\left(\begin{array}{ccccccccc}
4\lambda_{1} & \lambda_{5}+\lambda_{5}' & \sqrt{2}\lambda_{1} & \frac{\lambda_{5}}{\sqrt{2}} & \sqrt{2}\lambda_{1} & \frac{\lambda_{5}}{\sqrt{2}} & \frac{\lambda_{6}}{\sqrt{2}} & \frac{\lambda_{6}}{\sqrt{2}} & 2\lambda_{6}'\\
* & 4\lambda_{2} & \frac{\lambda_{5}}{\sqrt{2}} & \sqrt{2}\lambda_{2} & \frac{\lambda_{5}}{\sqrt{2}} & \sqrt{2}\lambda_{2} & \frac{\lambda_{7}}{\sqrt{2}} & \frac{\lambda_{7}}{\sqrt{2}} & 2\lambda_{7}'\\
* & * & 3\lambda_{1}  & \frac{1}{2}(\lambda_{5}+\lambda_{5}')& \lambda _1 & \frac{1}{2}(\lambda_{5}+\lambda_{5}') & \frac{\lambda_{6}}{2} & \frac{\lambda_{6}}{2} & \sqrt{2}\lambda_{6}'\\
* & * & * & 3\lambda_{2} & \frac{1}{2}(\lambda_{5}+\lambda_{5}') & \lambda_{2} & \frac{\lambda_{7}}{2} & \frac{\lambda_{7}}{2} & \sqrt{2}\lambda_{7}'\\
* & * & * & * & 3\lambda_{1} & \frac{1}{2}(\lambda_{5}+\lambda_{5}') & \frac{\lambda_{6}}{2} & \frac{\lambda_{6}}{2} & \sqrt{2}\lambda_{6}'\\
* & * & * & * & * & 3\lambda_{2} & \frac{\lambda_{7}}{2} & \frac{\lambda_{7}}{2} & \sqrt{2}\lambda_{7}'\\
* & * & * & * & * & * &3 \lambda_{3} & \lambda_{3}& \sqrt{2}\lambda_{8} \\
* & * & * & * & * & * & * & 3\lambda_{3} & \sqrt{2}\lambda_{8}\\
* & * & * & * & * & * & * & * & 24\lambda_{4}
\end{array}\right).
\end{eqnarray}
Its analytical eigenvalues are $2\lambda_{1}$, $2\lambda_{2}$, $2\lambda_{3}$ and

\begin{equation}
a_{\pm}=\lambda_{1}+\lambda_{2}\pm\sqrt{(\lambda_{1}-\lambda_{2})^{2}+(\lambda_{5}')^{2}}.
\end{equation}
The remaining four eigenvalues $b_{j}$, $j=1,2,3,4$ comes from the
solutions of a quartic polynomial equation that is not included here,
nevertheless it gives two double degenerate eigenvalues that according
to Eq. (\ref{eq:bound}) satisfy

\begin{equation}
\lambda_{1}+\lambda_{2}+\frac{2}{3}\lambda_{3}+4\lambda_{4}\leq16\pi.\label{eq:quartic polynomic condition}
\end{equation}

Finally, in the basis $(h_{1}z_{1},h_{2}z_{2},h_{3}z_{3})$ we obtain:

\begin{equation}
\mathcal{M}_{3}=\left(\begin{array}{ccc}
2\lambda_{1} & 0 & 0\\
0 & 2\lambda_{2} & 0\\
0 & 0 & 2\lambda_{3}
\end{array}\right).
\end{equation}

Thus, taking the unitarity condition from Eq. (\ref{eq:bound}), we find the bound

\begin{equation}
e_{1},\left|e_{2}\right|,\left|f_{\pm}\right|,\left|f_{1}\right|,2\lambda_{1},2\lambda_{2},2\lambda_{3},\left|a_{\pm}\right|,\left|b_{j}\right|\leq8\pi.\label{eq:inequnitarity}
\end{equation}

\subsection{Vacuum stability}



The stability condition in the strong sense of \cite{key-8} can be implemented by the definition
of the $\underline{K}$ like matrices

\begin{align}
\underline{K} & =\left(\begin{array}{cc}
\phi_{1}^{\dagger}\phi_{1} & \phi_{2}^{\dagger}\phi_{1}\\
\phi_{1}^{\dagger}\phi_{2} & \phi_{2}^{\dagger}\phi_{2}
\end{array}\right),\;\underline{L}=\left(\begin{array}{cc}
\chi_{0}^{*}\chi_{0} & 0
\\
0 & \sigma_{0}^{*}\sigma_{0}
\end{array}\right),\;\underline{M}=\left(\begin{array}{cc}
\phi_{1}^{\dagger}\phi_{1} & 0\\
0 & \chi_{0}^{*}\chi_{0}
\end{array}\right)\nonumber \\
\\
\underline{N} & =\left(\begin{array}{cc}
\phi_{1}^{\dagger}\phi_{1} & 0\\
0 & \sigma_{0}^{*}\sigma_{0}
\end{array}\right),\;\underline{P}=\left(\begin{array}{cc}
\phi_{2}^{\dagger}\phi_{2} & 0\\
0 & \chi_{0}^{*}\chi_{0}
\end{array}\right),\;\underline{Q}=\left(\begin{array}{cc}
\phi_{2}^{\dagger}\phi_{2} & 0\\
0 & \sigma_{0}^{*}\sigma_{0}
\end{array}\right).\nonumber
\end{align}
The above matrices can be decomposed in terms of the
Pauli matrices. For example, the components of $\underline{K}$ can be written as:

\begin{equation}
\underline{K}_{ij}=\frac{1}{2}\left(K_{0}\delta_{ij}+K_{a}\sigma_{ij}^{a}\right),
\end{equation}
where $K_{0}=\phi_{i}^{\dagger}\phi_{i}$ and $K_{a}=(\phi_{i}^{\dagger}\phi_{j})\sigma_{ij}^{a}$
for $i,j\in\left\{ 1,2\right\} $ and $a=1,2,3$. Extending the above decomposition to all matrices, we obtain the following components:

\begin{equation}
\underline{K}:\begin{cases}
\phi_{1}^{\dagger}\phi_{1}= & (K_{0}+K_{3})/2,\quad\phi_{1}^{\dagger}\phi_{2}=(K_{1}+iK_{2})/2,\\
\phi_{2}^{\dagger}\phi_{2}= & (K_{0}-K_{3})/2,\quad\phi_{2}^{\dagger}\phi_{1}=(K_{1}-iK_{2})/2,
\end{cases}
\end{equation}

\begin{equation}
\underline{L}:\begin{cases}
\chi_{0}^{*}\chi_{0}= & (L_{0}+L_{3})/2,
\\
\sigma_{0}^{*}\sigma_{0}= & (L_{0}-L_{3})/2,
\end{cases}
\end{equation}

\begin{equation}
\underline{M}:\begin{cases}
\phi_{1}^{\dagger}\phi_{1}= & (M_{0}+M_{3})/2\\
\chi_{0}^{*}\chi_{0}= & (M_{0}-M_{3})/2
\end{cases},\;\underline{N}:\begin{cases}
\phi_{1}^{\dagger}\phi_{1}= & (N_{0}+N_{3})/2,\\
\sigma_{0}^{*}\sigma_{0}= & (N_{0}-N_{3})/2,
\end{cases}
\end{equation}

\begin{equation}
\underline{P}:\begin{cases}
\phi_{2}^{\dagger}\phi_{2}= & (P_{0}+P_{3})/2\\
\chi_{0}^{*}\chi_{0}= & (P_{0}-P_{3})/2
\end{cases},\;\underline{Q}:\begin{cases}
\phi_{2}^{\dagger}\phi_{2}= & (Q_{0}+Q_{3})/2,\\
\sigma_{0}^{*}\sigma_{0}= & (Q_{0}-Q_{3})/2.
\end{cases}
\end{equation}

Thus, the potential in Eq. (\ref{eq:quartic potential}) become:

\begin{equation}
V_{4}=\sum_{r}V_{4r},\quad r=K,L,M,N,P,Q,
\end{equation}
which can be written as \cite{key-8} :

\begin{equation}
V_{4r}=\eta_{r00}r_{0}^{2}+2r_{0}\eta_{r}r_{a}+r_{a}E_{r}r_{b},
\end{equation}
with

\begin{equation}
\begin{array}{cc}
\eta_{K00}=\frac{1}{4}(\lambda_{1}+\lambda_{2}+\lambda_{5}), & \eta_{L00}=\frac{1}{4}(\lambda_{3}+\lambda_{4}+\lambda_{8}),\\
\eta_{M00}=\frac{1}{4}\lambda_{6},\;\eta_{N00}=\frac{1}{4}\lambda_{6}', & \eta_{P00}=\frac{1}{4}\lambda_{7},\;\eta_{Q00}=\frac{1}{4}\lambda_{7}',
\end{array}
\end{equation}

\begin{equation}
\begin{array}{cc}
\boldsymbol{\eta}_{K}=\frac{1}{4}\left(\begin{array}{c}
0\\
0\\
\lambda_{1}-\lambda_{2}
\end{array}\right), & \boldsymbol{\eta}_{L}=\frac{1}{4}\left(\begin{array}{c}
0\\
0\\
\lambda_{3}-\lambda_{4}
\end{array}\right),\\
\boldsymbol{\eta}_{r}=0,\quad r=M,N,P,Q
\end{array}\label{eq:Etas}
\end{equation}

\begin{equation}
E_{K}=\frac{1}{4}\left(\begin{array}{ccc}
\lambda_{5}' & 0 & 0\\
0 & \lambda_{5}' & 0\\
0 & 0 & \lambda_{1}+\lambda_{2}-\lambda_{5}
\end{array}\right),\quad E_{L}=\frac{1}{4}\left(\begin{array}{ccc}
0 & 0 & 0\\
0 & 
0 & 0\\
0 & 0 & \lambda_{3}+\lambda_{4}-\lambda_{8}
\end{array}\right).
\end{equation}

The strong stability condition requires that $f_{r}(u_{i})>0$ for all $u_{i}$ in a set $I=\left\{ u_{1},...,u_{n}\right\}$ \cite{key-8}, where the function $f_{r}(u)$ is defined as

\begin{equation}
f_{r}(u)=u+\eta_{r00}-\boldsymbol{\eta}_{r}^{T}(E_{r}-u)^{-1}\boldsymbol{\eta}_{r},
\end{equation}
and its derivative:

\begin{equation}
f_{r}'(u)=1-\boldsymbol{\eta}_{r}^{T}(E_{r}-u)^{-2}\boldsymbol{\eta}_{r}.
\end{equation}
For example, if $r=K$, we obtain:

\[
f_{K}(u)=u+\frac{1}{4}(\lambda_{1}+\lambda_{2}+\lambda_{5})-\frac{(\lambda_{1}-\lambda_{2})^{2}}{4(\lambda_{1}+\lambda_{2}-\lambda_{5}-4u)},
\] \\
\[
f_{K}'(u)=1-\frac{(\lambda_{1}-\lambda_{2})^{2}}{(\lambda_{1}+\lambda_{2}-\lambda_{5}-4u)^{2}}.
\]
The roots of the derivative are:

\begin{align*}
f_{K}'(u) & =0\longrightarrow
\begin{cases}
 u_{1}=\frac{1}{4}(2\lambda_{2}-\lambda_{5}), \\
 u_{2}=\frac{1}{4}(2\lambda_{1}-\lambda_{5}).
 \end{cases}
\end{align*}

We evaluate the stability condition $f_{K}(u_{i})>0$ in the set $I=\left\{0, u_{1}, u_{2}, \mu \right\}$, where $\mu = \lambda _5'/4$ corresponds to the doubly degenerated eigenvalue of the $E_{K}$ matrix such that $f_{K}(\mu )$ is finite
and $f_{K}'(\mu )\geq0$, obtaining:


\begin{equation}
f_{K}(0)>0 \ \ \ \rightarrow \ \ \ 4\lambda_{1}\lambda_{2}>\lambda_{5}^{2}. \label{stab-1}
\end{equation}


\begin{align}
f_{K}(u_{1,2})>0 \ \ \ \rightarrow\ \ \ \lambda_{1}+\lambda_{2}>0,
\end{align}


\begin{equation}
f_{K}(\mu )>0\ \ \ \rightarrow\ \ \ 4\lambda_{1}\lambda_{2}>(\lambda_{5}+\lambda_{5}')^{2}.
\end{equation}

With an identical procedure for $f_{L}(u_{i})>0$, we obtain:

\begin{align}
4\lambda_{3}\lambda_{4}>\lambda_{8}^{2},\\ \nonumber \\
\lambda_{3}+\lambda_{4}>0,\label{stab-5}
\end{align}

The matrices $E_{r}$ for $r=M,N,P,Q$ are reduced trivially to one
element (eigenvalue) due to (\ref{eq:Etas})

\begin{align}
V_{4M} & =M_{a}E_{Mab}M_{b},\; V_{4N}=N_{a}E_{Nab}N_{b},\nonumber \\
V_{4P} & =P_{a}E_{Pab}P_{b},\; V_{4Q}=Q_{a}E_{Qab}Q_{b},
\end{align}

obtaining from the condition $f_{r}(0)>0$

\begin{equation}
\lambda_{6}>0,\;\lambda_{6}'>0,\;\lambda_{7}>0,\;\lambda_{7}'>0.\label{eq:ineq3}
\end{equation}



\subsection{Combined constraints\label{constraint}}

The unitarity conditions in Eqs. (\ref{eq:quartic polynomic condition}) and  (\ref{eq:inequnitarity})
and the stability ones (\ref{stab-1}-\ref{stab-5}) and (\ref{eq:ineq3}),
can be combined in order to obtain a more suitable parameter space.
In this way, the final combined conditions are

\begin{align}
0<\lambda_{1,2,3}<4\pi,\label{eq:lambda1}\\
\lambda_{4}>0,\\
\lambda_{5}\leq8\pi,\\
\lambda_{5}+2\lambda_{5}'\leq8\pi\\
\lambda_{1}+\lambda_{2}+\frac{2}{3}\lambda_{3}+4\lambda_{4}\leq16\pi,\\
\left|\lambda_{5}\right|<2\sqrt{\lambda_{1}\lambda_{2}},\label{eq:lambda5} \\
\left|\lambda_{5}+\lambda_{5}'\right|<2\sqrt{\lambda_{1}\lambda_{2}},\label{eq:lmabda 5 lambda prime}\\
\left|\lambda_{8}\right|<2\sqrt{\lambda_{3}\lambda_{4}},\\
\lambda_{6,7}>0,\\
\lambda_{6,7}'>0.
\label{lambda-sigma}
\end{align}




\section{Diphoton Higgs decay\label{sec:diphoton}}

In the SM, the decay of the Higgs boson to diphoton is mediated by fermions and charged vector boson loops. In the $U(1)'$ model, there is an additional contribution due to the charged Higgs boson loops, obtaining the total diphoton Higgs width \cite{djouadi}

\begin{eqnarray}
\Gamma (h_0\rightarrow \gamma \gamma)=\frac{\alpha ^2 M_{h_0}^3}{256 \pi ^3 \upsilon ^2}\left|F_1(\tau _W)+\sum _f N_{cf}Q_f^2F_{1/2}(\tau _f)+g_{H^{\pm}}F_0(\tau _{H^{\pm }}) \right|^2,
\label{diphoton-width}
\end{eqnarray}
where $N_{cf}$ and $Q_f$ are the color and electric charge factors, respectively, and:

\begin{eqnarray}
\tau _{a}=\frac{4M_{a}^2}{M_{h_0}^2},
\end{eqnarray}
for $a=W, f$ and $H^{\pm}$. The loop factors are:

\begin{eqnarray}
F_1&=&2+3\tau +3\tau (2-\tau )f(\tau ), \nonumber \\
F_{1/2}&=&-2\tau [1+(1-\tau )f(\tau )], \nonumber \\
F_0&=&\tau [1-\tau f(\tau )],
\end{eqnarray}
with:

\begin{eqnarray}
f(\tau )=
\begin{cases}
\left[\sin ^{-1}(1/\sqrt{\tau })\right]^2, & \tau \ge 1 \\
-\frac{1}{4}\left[\ln \left(\eta _+/\eta _- \right)-i\pi \right]^2, & \tau < 1
\end{cases}
\end{eqnarray}
where $\eta _{\pm}=1\pm \sqrt{1-\tau }$. The charged Higgs coupling $g_{H^{\pm}}$ is given by Eq. (\ref{trilineal-coup}). 


On the other hand, the theoretical signal strenght is defined as the ratio between the $h_0 \rightarrow \gamma \gamma$ branching decay of the $U(1)'$ model and the SM prediction:

\begin{eqnarray}
R_{\gamma \gamma}=\frac{Br(h_0 \rightarrow \gamma \gamma)}{Br(h_0 \rightarrow \gamma \gamma)^{SM}}.
\end{eqnarray}
We identify two scenarios according to the mass of the DM candidate of the model:
\begin{itemize}
\item[-] \textit{Scenario I:} If $M_{\sigma } > M_{h_{0}}/2 \approx 63$ GeV, the decay of the Higgs boson to DM pair is kinematically forbidden. By assuming that the final states of the Higgs boson decay are of SM nature, then the signal strenght can be approximated as

\begin{eqnarray}
R_{\gamma \gamma}\approx \frac{\Gamma(h_0 \rightarrow \gamma \gamma)}{\Gamma(h_0 \rightarrow \gamma \gamma)^{SM}}.
\label{signal-no-dm}
\end{eqnarray}
where the width of the SM is the same as Eq. (\ref{diphoton-width}) but without the $F_0$ factor.

\item[-] \textit{Scenario II:} If $M_{\sigma } \leq M_{h_{0}}/2 \approx 63$ GeV, the decay of the Higgs boson to DM pair is allowed. In this case, the total decay width can be separated in decays to SM particles and decay to DM particles. Thus, we obtain:

\begin{eqnarray}
R_{\gamma \gamma}\approx \frac{\Gamma(h_0 \rightarrow \gamma \gamma)\times \Gamma_{h_{0}}^{SM}}{\Gamma(h_0 \rightarrow \gamma \gamma)^{SM}\times \left[\Gamma_{h_{0}}^{SM}+\Gamma(h_0 \rightarrow \sigma \sigma ^{*})\right]}.
\label{signal-dm}
\end{eqnarray}
where $\Gamma_{h_{0}}^{SM}$ is the total decay width of the SM Higgs boson, while the  width to DM pair is:

\begin{eqnarray}
\Gamma(h_0 \rightarrow \sigma \sigma ^{*})=\frac{g_{\sigma }^2}{16\pi M_{h_{0}}}\sqrt{1-\frac{4M_{\sigma _0}^2}{M_{h_{0}}^2}},
\end{eqnarray}
and the coupling $g_{\sigma}$ is given in (\ref{trilineal-coup}). 


\end{itemize}

According to the trilineal couplings $g_{H^{\pm}}$ and $g_{\sigma }$ in Eq. (\ref{trilineal-coup}), the signal strenght depends on the Higgs couplings $\lambda _{1,2,5}$ and $\lambda '_{5,6,7}$ from the quartic terms of the potential, which we rewrite as follow: 

\begin{eqnarray}
V_{4}&=&\lambda _1 \left| \phi _1 \right|^4+\lambda _2 \left| \phi _2 \right|^4+\lambda _5 \left| \phi _1 \right|^2\left| \phi _2 \right|^2+\lambda '_5 \left| \phi _1^{\dagger} \phi _2 \right|^2\nonumber \\
&+&\lambda '_6\left| \phi _1 \right|^2\left| \sigma _0\right|^2+\lambda '_7 \left| \phi _2 \right|^2\left| \sigma _0\right|^2.
\label{higgs-pot-cuartic}
\end{eqnarray}

In order to evaluate the constraints from the diphoton decay, we impose some assumptions of our space of parameters. First, since the scalar couplings $\lambda _1$ and $\lambda _2$ show the same theoretical constraints as observed in the subsection \ref{constraint}, we can assume only one characteristic diagonal coupling constant $\lambda _D=\lambda _1=\lambda _2$. Thus, we suppose that each doublet $\phi _1$ and $\phi _2$ shows the same self-interaction separately. On the other hand, there are two types of mixing couplings between both doublets, distinguished by the coupling constants $\lambda _5$ and $\lambda _5'$. In this case, we also assume one characteristic coupling for the mixing interaction between the scalar doublets. Thus, we choose $\lambda _5 = \lambda _5'$ as the mixing coupling between $\phi _1$ and $\phi _2$. For the interactions of the scalar doublets with the scalar singlet $\sigma _0$, we assume that $\lambda '=\lambda _6'=\lambda _7'$. For the numerical analysis, it will be convenient to define the ratio

\begin{eqnarray}
r_{\lambda }=\frac{\lambda _{5}}{\lambda _D}.
\label{coupling-ratio}
\end{eqnarray}

With the above parametrization, the constraints from Eqs. (\ref{eq:lambda1}), (\ref{eq:lmabda 5 lambda prime}) and (\ref{lambda-sigma}) become:

\begin{eqnarray}
0<& \lambda _D =\lambda _1=\lambda _2& <4\pi \nonumber \\
-1<&r_{\lambda}=\frac{\lambda _{5}}{\lambda _D}=\frac{\lambda '_{5}}{\lambda _D}&<1 \nonumber \\
&\lambda '=\lambda '_6=\lambda '_7&>0. 
\label{theoretical-constraint-reduce}
\end{eqnarray}

The charged and DM scalar coupling functions from Eqn. (\ref{trilineal-coup}) become:

\begin{eqnarray}
g_{H^{\pm}}&=&\upsilon \lambda _D\left(S_{2\beta }S_{(\alpha+\beta)}+r_{\lambda}C_{2\beta}C_{(\alpha+\beta)} \right) \nonumber \\
g_{\sigma }&=&\upsilon  \lambda 'C_{\alpha -\beta },
\label{trilineal-coup-2}
\end{eqnarray}
while the rotation angles and the mass of the Higgs boson $h_0$ in Eqs. (\ref{alpha-angle}) and (\ref{scalar-mass}) can be written as:

\begin{eqnarray}
\sin ^2{\alpha }&\approx& \sin ^2{\beta }\left[1+\frac{4\upsilon ^2}{M_{H^{\pm }}^2}C_{2\beta }C_{\beta }^2\lambda _D(1-r_{\lambda })\right] \label{angle-relations-x} \\
M_{h_0}^2 &\approx& 2\upsilon ^2\lambda _D\left[1-\frac{1}{2}S_{2\beta }^2(1-r_{\lambda })\right],
\label{higgs-mass-2}
\end{eqnarray}
where the mass $M_{H^{\pm }}^2$ is proportional to the large scale $f_2\upsilon _{\chi} $. However, since at dominant order the spectrum is degenerated, it could be replaced by any of the other two neutral Higgs bosons, according to Eq. (\ref{scalar-mass}).  
 
For the numerical analysis, we use the experimental data of the diphoton signal strenght $R_{\gamma \gamma}=1.55^{+0.33}_{-0.28}$ obtained by ATLAS \cite{atlas-gama} and $R_{\gamma \gamma}=1.54^{+0.46}_{-0.42}$ at CMS \cite{cms-gama} for $M_{h_{0}}=125.5$ GeV.

\subsection{Scenario I}

The set of parameters involved in the diphoton Higgs decay in the scenario I is \\
$(T_{\beta }, T_{\alpha }, \lambda _D, r_{\lambda }, M_{H^{\pm}})$. However, the two relations shown in Eqs. (\ref{angle-relations-x}) and (\ref{higgs-mass-2}) impose model-dependent constraints that reduces the number of free parameters. If we initially take into account only the relation between angles from (\ref{angle-relations-x}), then the signal strenght in (\ref{signal-no-dm}) depends on $(T_{\beta}, \lambda _D, r_{\lambda }, M_{H^{\pm}})$. 

\begin{figure}[tbh]
\centering
\includegraphics[width=5cm,height=6cm,angle=0]{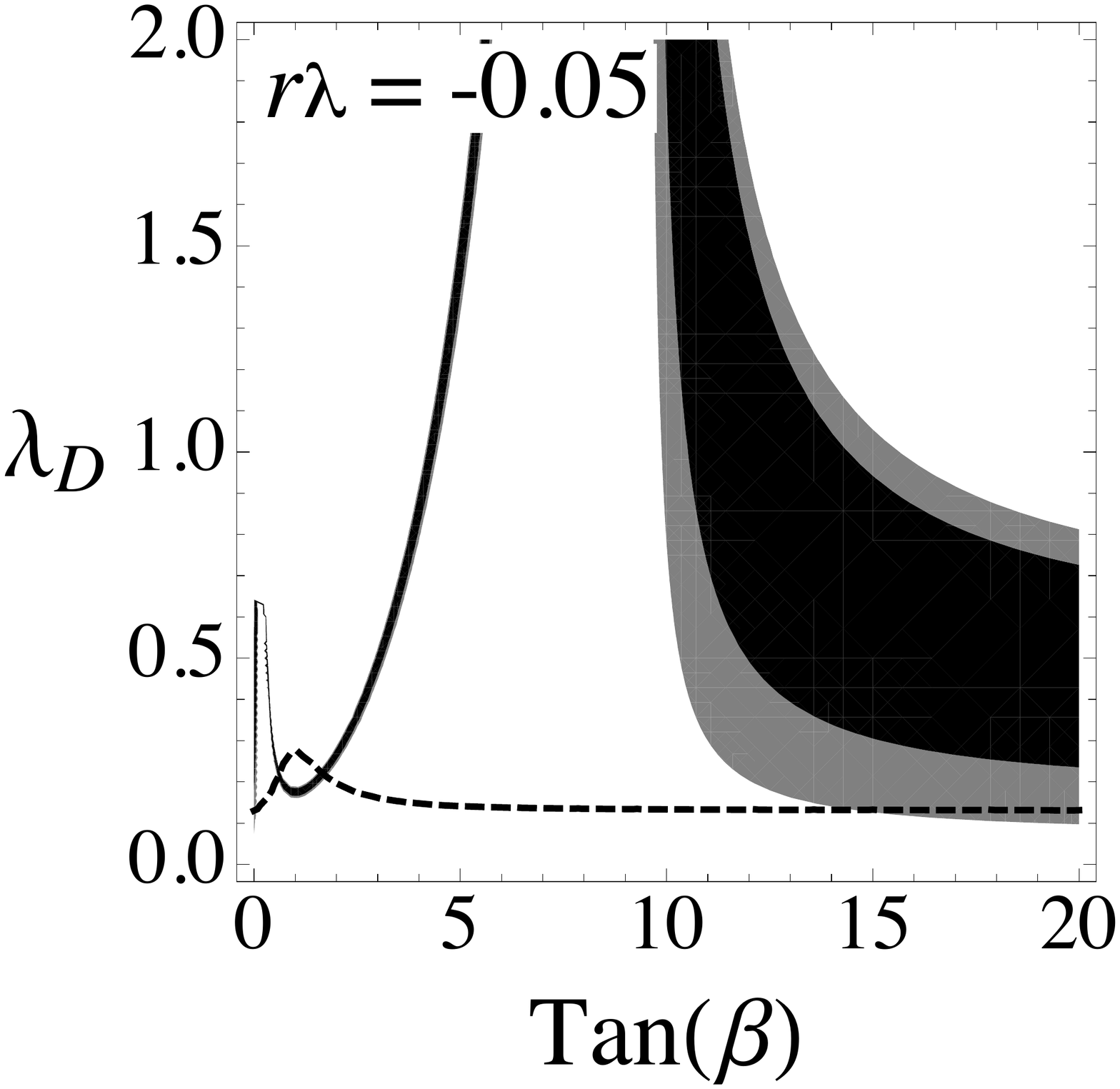}
\includegraphics[width=5cm,height=6cm,angle=0]{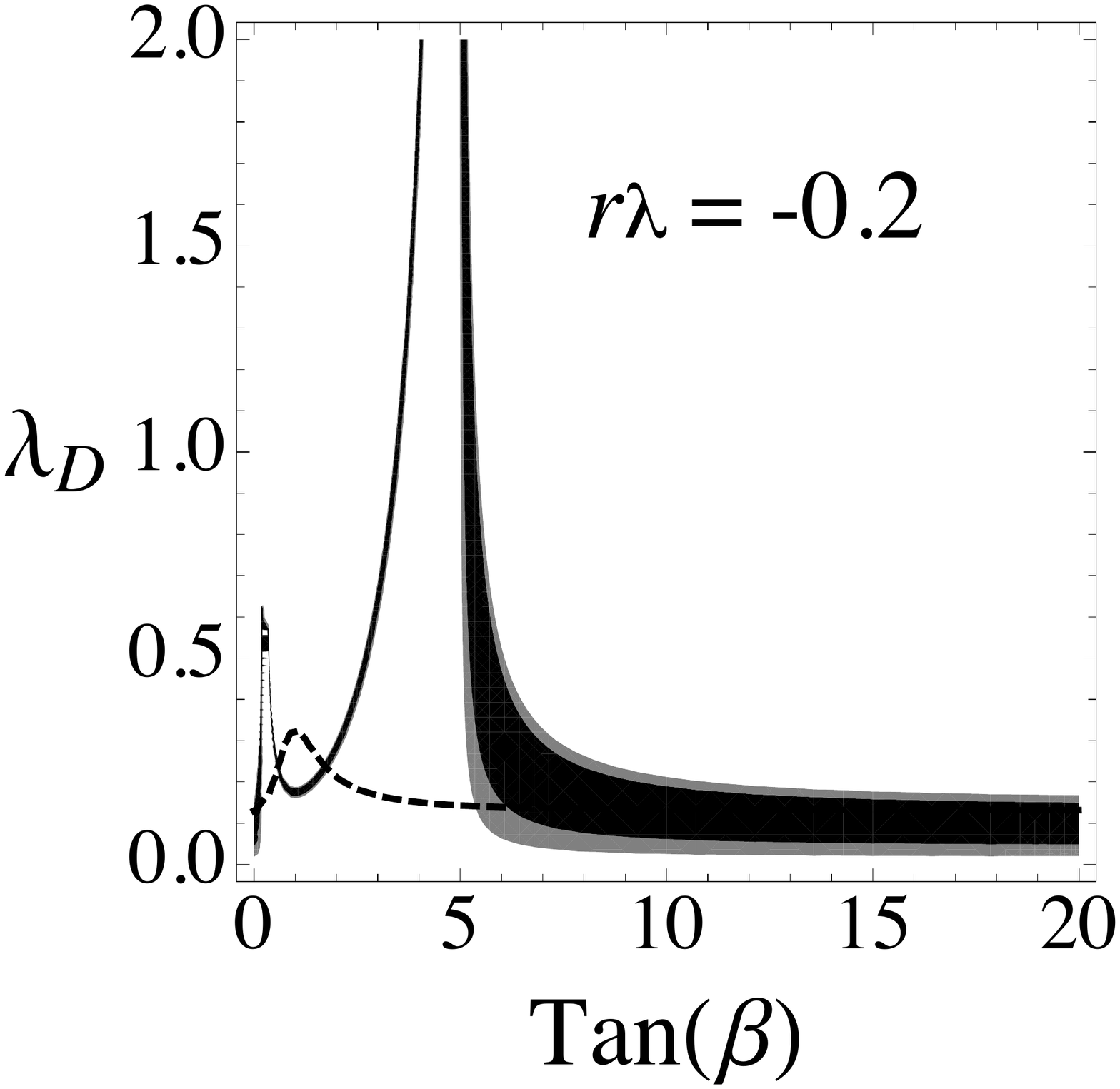}
\includegraphics[width=5cm,height=6cm,angle=0]{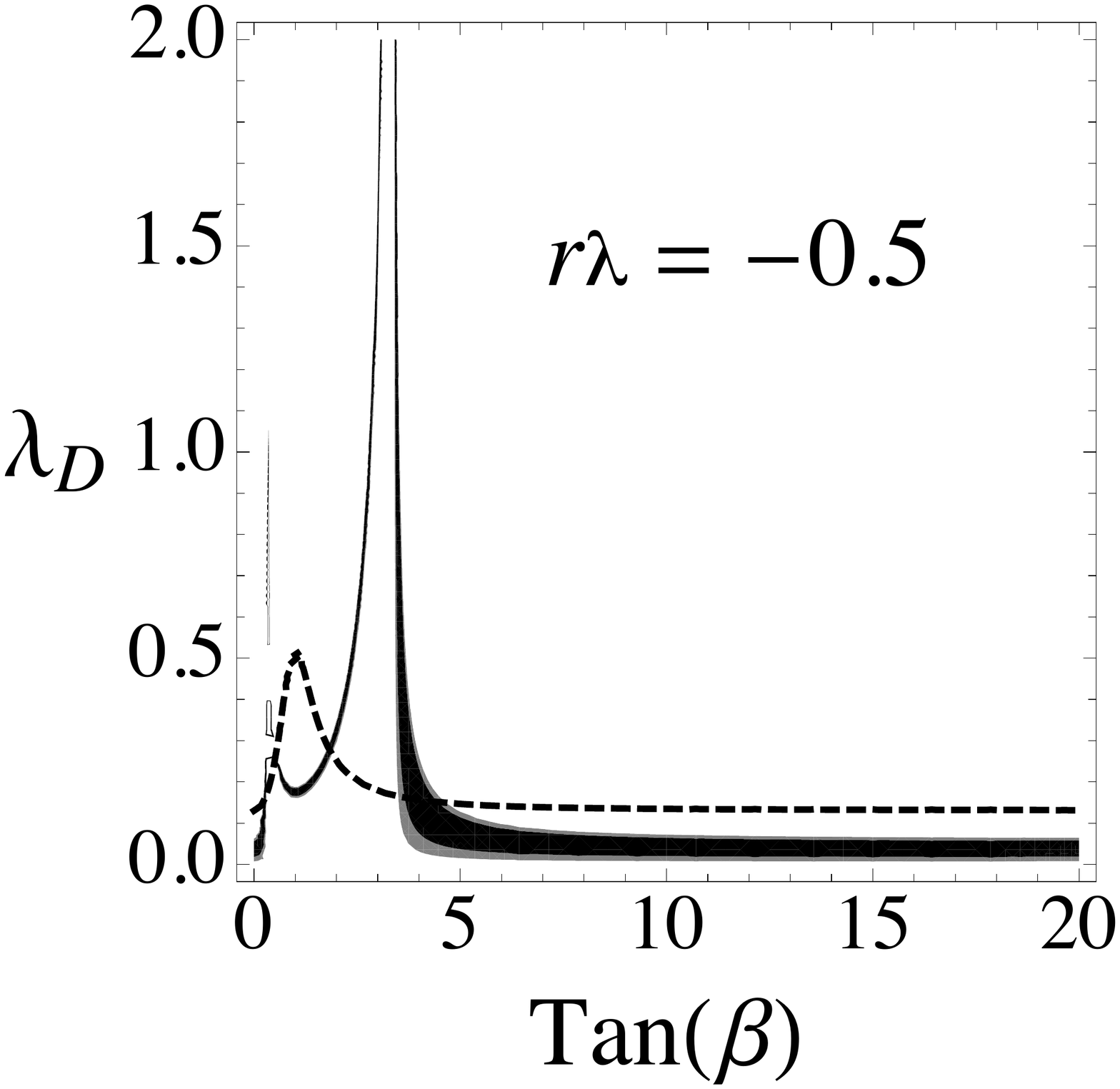}\\
\includegraphics[width=5cm,height=6cm,angle=0]{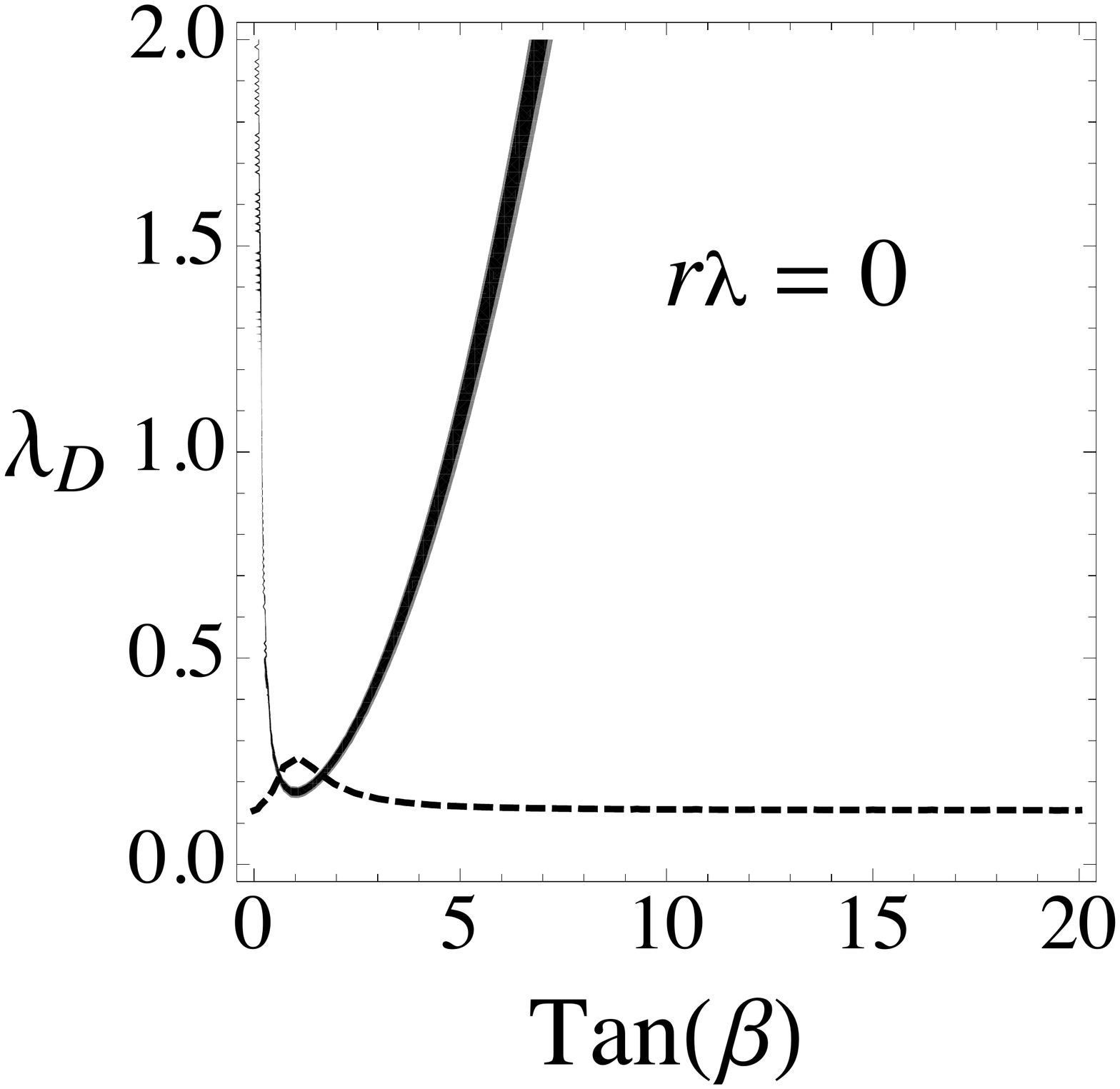}
\includegraphics[width=5cm,height=6cm,angle=0]{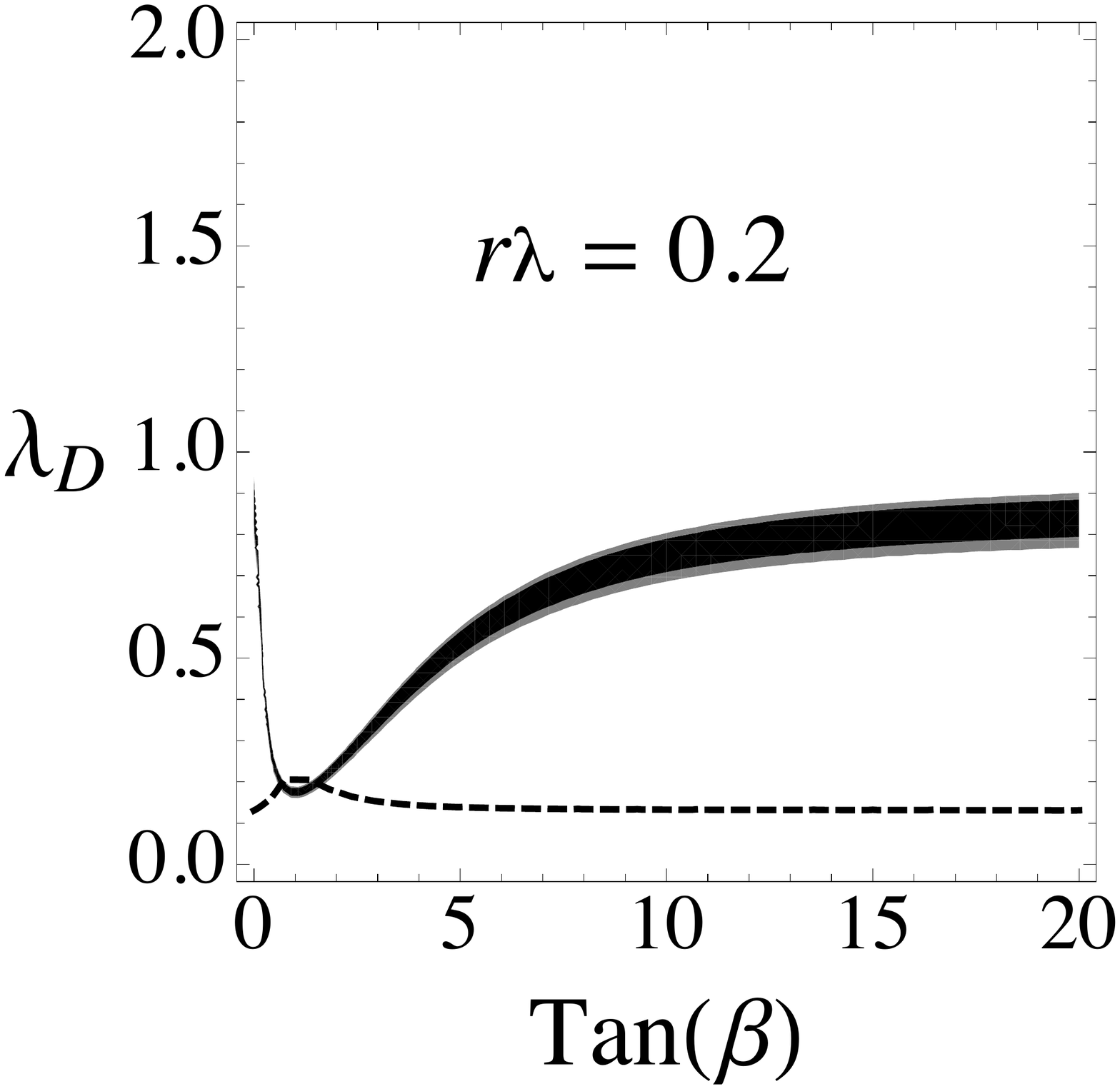}
\includegraphics[width=5cm,height=6cm,angle=0]{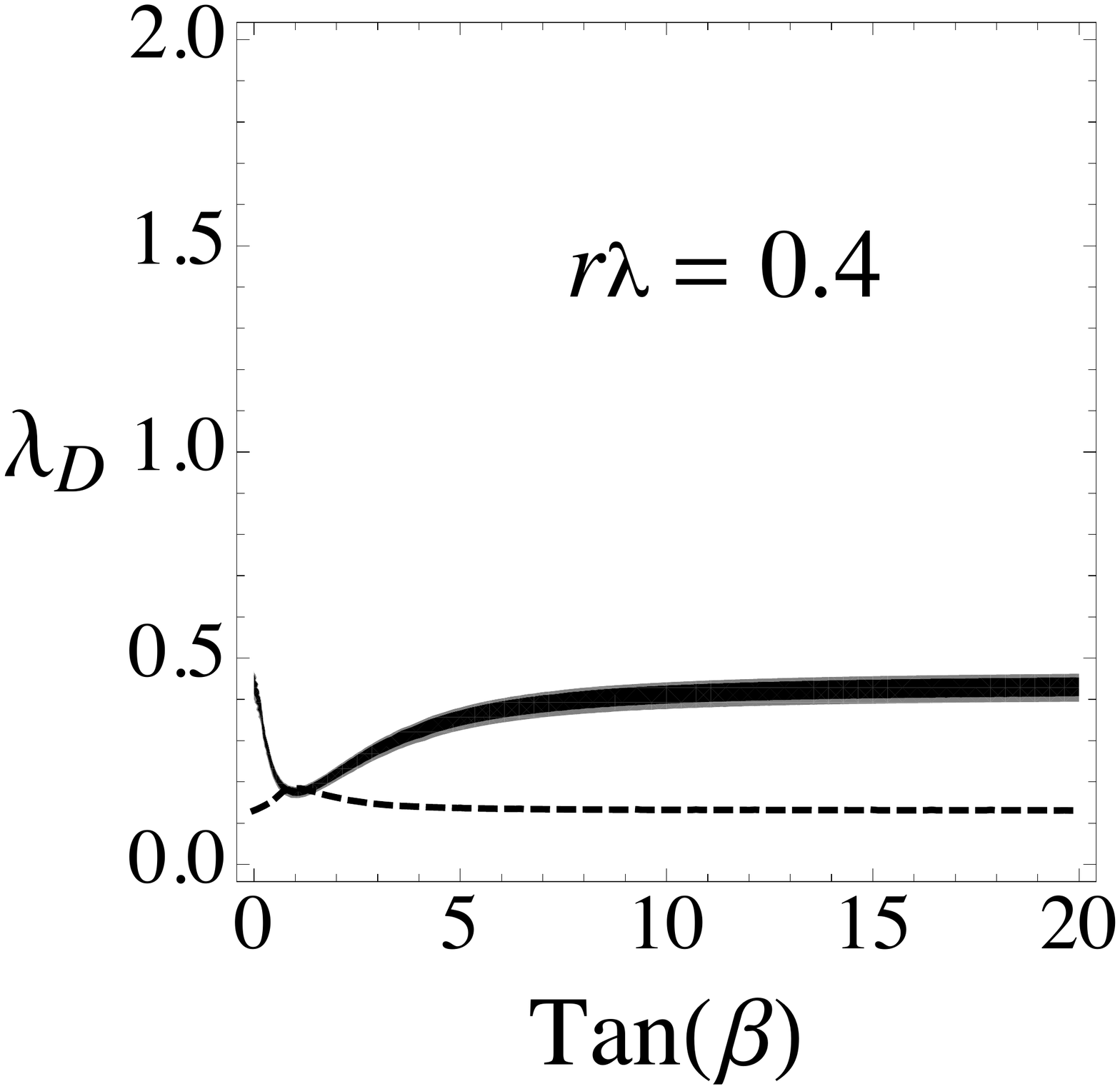}
\caption{Allowed regions in the $(T_{\beta },\lambda _D)$ plane, compatible with the diphoton Higgs decay limits at ATLAS (black region) and CMS (gray region), for six values of the coupling ratio $r_{\lambda }$. The charged Higgs mass is fixed to be $M_{H^{\pm}}=300$ GeV. The dashed line corresponds to $\lambda _D$ as function of $T_{\beta }$ obtained from Eq. (\ref{higgs-mass-2}) for $M_{h_0}=125.5$ GeV.}
\label{fig1}
\end{figure}

\begin{figure}[tbh]
\centering
\includegraphics[width=5cm,height=6cm,angle=0]{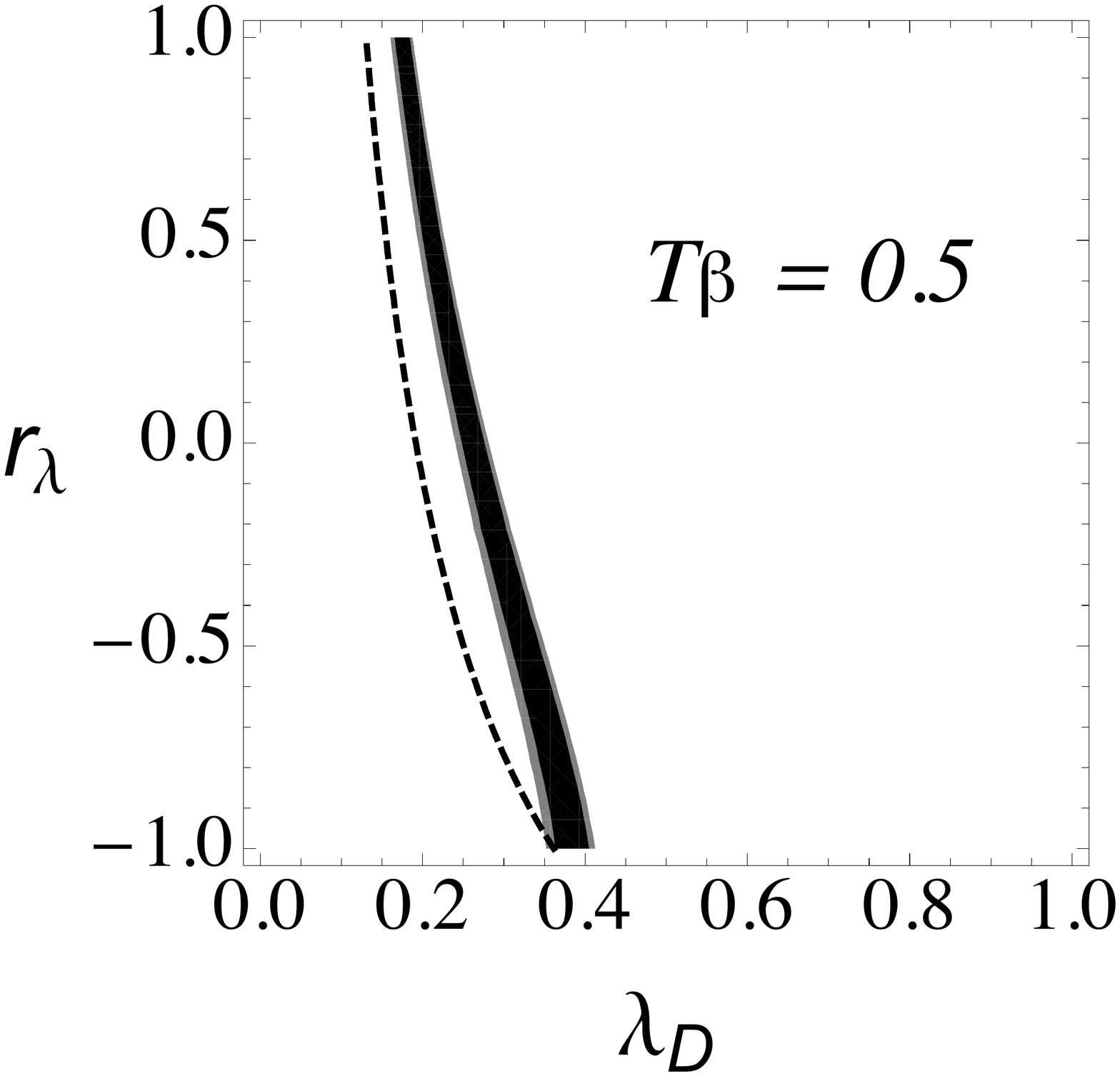}
\includegraphics[width=5cm,height=6cm,angle=0]{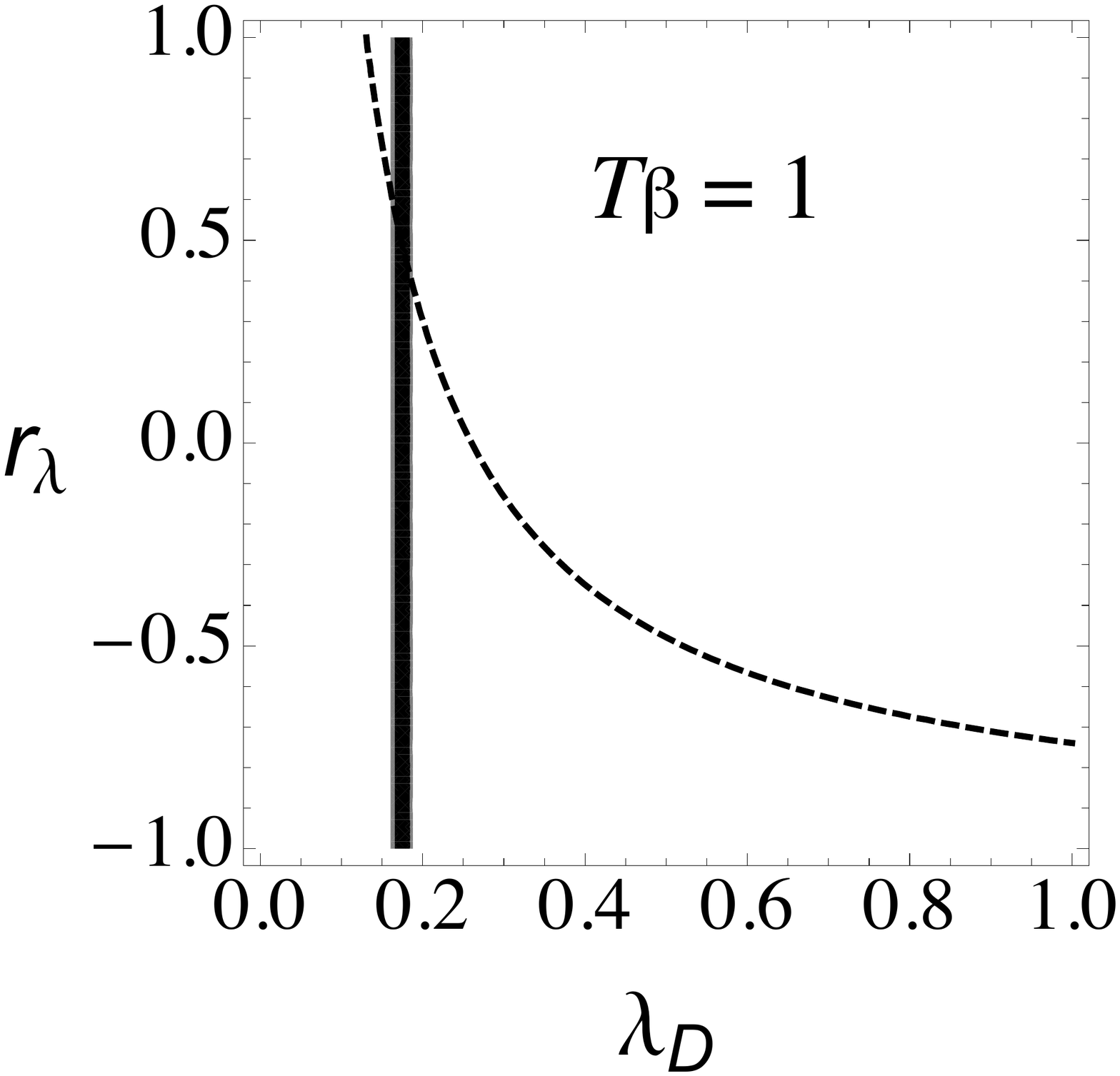}
\includegraphics[width=5cm,height=6cm,angle=0]{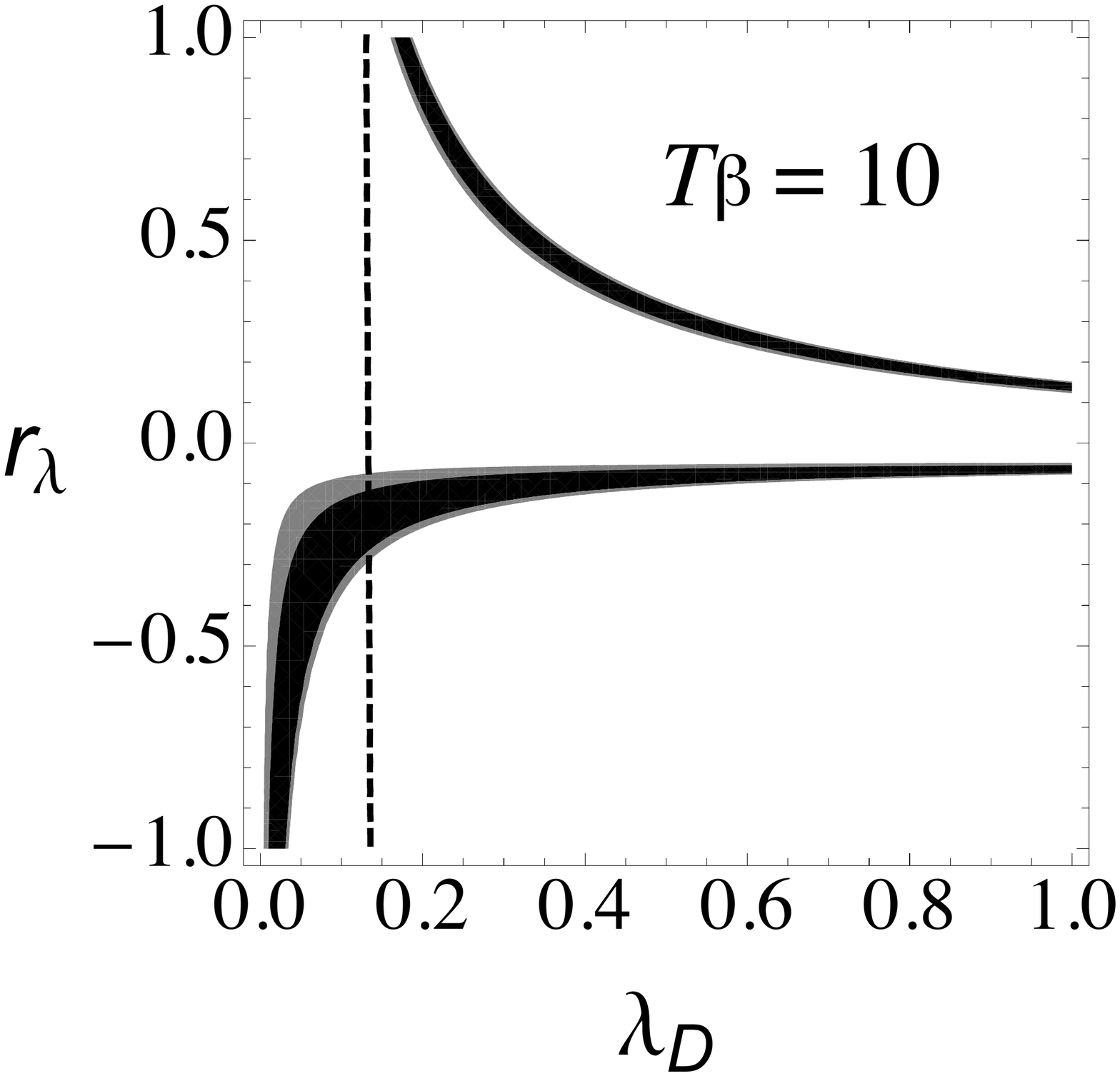}\vspace{-1cm}
\caption{Allowed regions in the $(\lambda _D,r_{\lambda })$ plane, compatible with the diphoton Higgs decay limits at ATLAS (black region) and CMS (gray region), for three values of $T_{\beta }$. The charged Higgs mass is fixed to be $M_{H^{\pm}}=300$ GeV. The dashed line corresponds to $r_{\lambda }$ as function of $\lambda _D$ from Eq. (\ref{higgs-mass-2}).}
\label{fig2}
\end{figure}

\begin{figure}[tbh]
\centering
\includegraphics[width=8cm,height=10cm,angle=0]{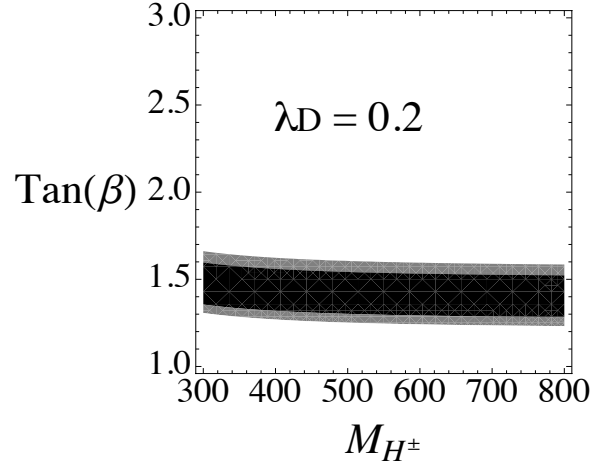}\vspace{-2cm}
\caption{Allowed regions in the $( M_{H^{\pm}},T_{\beta })$ plane, compatible with the diphoton Higgs decay limits at ATLAS (black region) and CMS (gray region), for $\lambda _{D}=0.2$. The coupling ratio is fixed to be $r_{\lambda }=0$.}
\label{fig3}
\end{figure}

Taking into account that the SM prediction for the diphoton branching is $Br(h_0 \rightarrow \gamma \gamma)^{SM}=2.28 \times 10^{-3}$, while its total width is $\Gamma _{h_{0}}=4.07 \times 10^{-3}$ GeV for a $125$ GeV SM Higgs boson \cite{data} we obtain the following constraints:

\begin{enumerate}
\item Fig. \ref{fig1} displays contour plots in the plane $T_{\beta }-\lambda _D$ for different values of the ratio $r_{\lambda }$. 
We fix the charged Higgs mass at $M_{H^{\pm}}=300$ GeV.  Since CMS reports a larger uncertainty range for the signal strenght, we can see broader regions than from ATLAS data in some plots. In particular, we see larger allowed intervals for small and negative $r_{\lambda }$ values. Thus, for example, if $r_{\lambda }=-0.05$ and $T_{\beta }=20$, we see allowed intervals as large as $\lambda _D=[0.1,0.8]$. If $T_{\beta }$ decreases near $10$, the $\lambda _D$ coupling can be larger than 2, but with small allowed intervals. We also see a very thin area for smaller $T_{\beta }$. A similar behavior is observed for other negative values, as shown in $r_{\lambda }=-0.2$ and $-0.5$ plots. For positive $r_{\lambda }$ values, the allowed bands change as shown, which exhibits smaller intervals.

On the other hand, if we impose the mass constraint from Eq. (\ref{higgs-mass-2}) with $M_{h_{0}}=125.5$ GeV, one of the three parameters $(T_{\beta }, \lambda _D, r_{\lambda })$ can be fixed as function of the others two. For example, the dashed lines in plots at Fig. \ref{fig1} correspond to $\lambda _D$ as function of $T_{\beta }$ for the six $r_{\lambda }$ values, obtaining very stringent constraints. First, we see that $\lambda _D$ takes values below $0.5$. Also, we see that although the plot with $r_{\lambda }=-0.05$ exhibits broader allowed areas, the mass constraint exclude most of the region. This case is compatible with CMS data for $T_{\beta} \geq 14$ and $\lambda _D=0.13$, while both, ATLAS and CMS points cut the dashed line for $T_{\beta }=1.7$ and $0.6$ at $\lambda _D$ around $0.2$. For $r_{\lambda }=-0.2$, the dashed line go through the experimental allowed regions for $T_{\beta } \geq 5.3$. Other values of $r_{\lambda }$ are practically ruled out except for puntual values or very narrow regions.

\item The above constraints are obtained for particular values of the ratio $r_{\lambda }$. In order to explore other solutions, we plot the allowed regions in the plane $\lambda _D-r_{\lambda }$ for different values of $T_{\beta }$, as shown in Fig. \ref{fig2}. The case with $T_{\beta }=0.5$ is completely excluded by the data if we take into account the mass constraint (dashed line). For $T_{\beta }=1$, we find a narrow allowed interval around $(\lambda _D,r_{\lambda})=(0.17,0.5)$. Only for large $T_{\beta }$ values, we obtain broader regions. For $T_{\beta }=10$, the experimental data are compatible with the mass constraint from Eq. (\ref{higgs-mass-2}) at $\lambda _D=0.13$ and $r_{\lambda }$ in the range between $-0.07$ and $0.3$.

\item In the above constraints, we fixed the heavy scale at $M_{H^{\pm}}=300$ GeV. However, the conclusions do not change significantly for larger values. To explore this, we show in Fig. \ref{fig3} the allowed region in the plane $M_{H^{\pm}}-T_{\beta }$, which exhibits a small variation with the mass.

\end{enumerate}

\subsection{Scenario II}

In this case, the signal strenght from Eq. (\ref{signal-dm}) depends on two additional parameters: $(\lambda ', M_{\sigma _0})$. According to the last plot in Fig. \ref{fig2} (for $T_{\beta }=10$), both constraints in Eqs. (\ref{angle-relations-x}) and (\ref{higgs-mass-2}) are compatible with the experimental data at $\lambda _D$ near $0.13$ and the central value $r_\lambda =-0.3$. Using these values, we show in Fig. \ref{fig4} the allowed points in the plane $(M_{\sigma _{0}}, \lambda ')$. We can see that the coupling with $\sigma _0$ takes small values at low masses and increases near the kinematic limit at $63$ GeV. For example, allowed range of $\lambda '$ between $0.02$ and $0.08$ is obtained for a $25$ GeV DM candidate, while for a $60$ GeV DM, the range become broader with values between $0.03$ and $0.15$. 

On the other hand, in order to evaluate the effects of invisible Higgs decays due to DM candidates on the diphoton signal strenght, we obtain again the contours in the plane $(\lambda _D, r_{\lambda })$ for different values of $\lambda '$. Fig. \ref{fig5} shows the allowed regions for $M_{\sigma _{0}}=60$ GeV and $T_{\beta }=10$. We see that large $\lambda '$ values produces large negative values on $r_{\lambda }$. The dashed line in the plots corresponds to the mass constraint from Eq. (\ref{higgs-mass-2}). Taking into account this constraint, we obtain the ranges $r_{\lambda }=[-0.1,-0.35]$, $[-0.2,-0.45]$ and $[-0.33,-0.6]$ for $\lambda '=0.05,$ $0.1$ and $0.15$, respectively. 

\section{Constraints for light dark matter\label{sec:DM}}

We can combine the data from DM direct and indirect detection experiments to explore additional implications and constraints of the model. Specifically, taking into account the constraints from the diphoton decay, we explore the allowed regions for light DM compatible with the observed DM relic density \cite{planck} and the limits from the LUX collaboration for the DM-nucleon cross section \cite{lux}. We use the micrOMEGAs software \cite{micromegas}.

\subsection{Relic density}

\begin{figure}[tbh]
\centering
\includegraphics[width=7cm,height=9cm,angle=0]{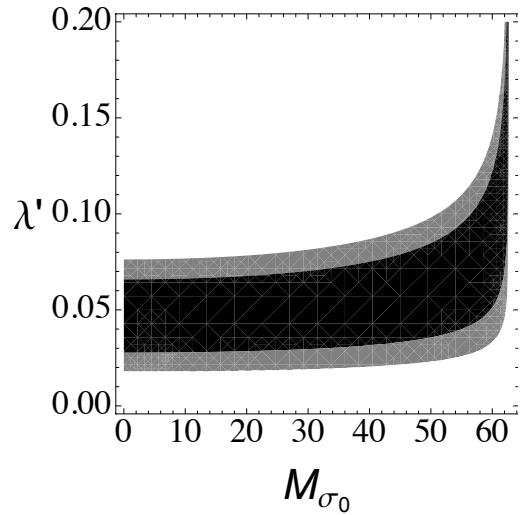}\vspace{-1cm}
\caption{Allowed regions in the $(M_{\sigma _{0}}, \lambda ')$ plane, compatible with the diphoton Higgs decay limits at ATLAS (black region) and CMS (gray region), for $\lambda _{D}=0.13$. The other parameters are fixed to be: $T_{\beta }=10$, $M_{H^{\pm }}=300$ GeV and $r_{\lambda }=-0.3$.}
\label{fig4}
\end{figure}

\begin{figure}[tbh]
\centering
\includegraphics[width=5cm,height=6.5cm,angle=0]{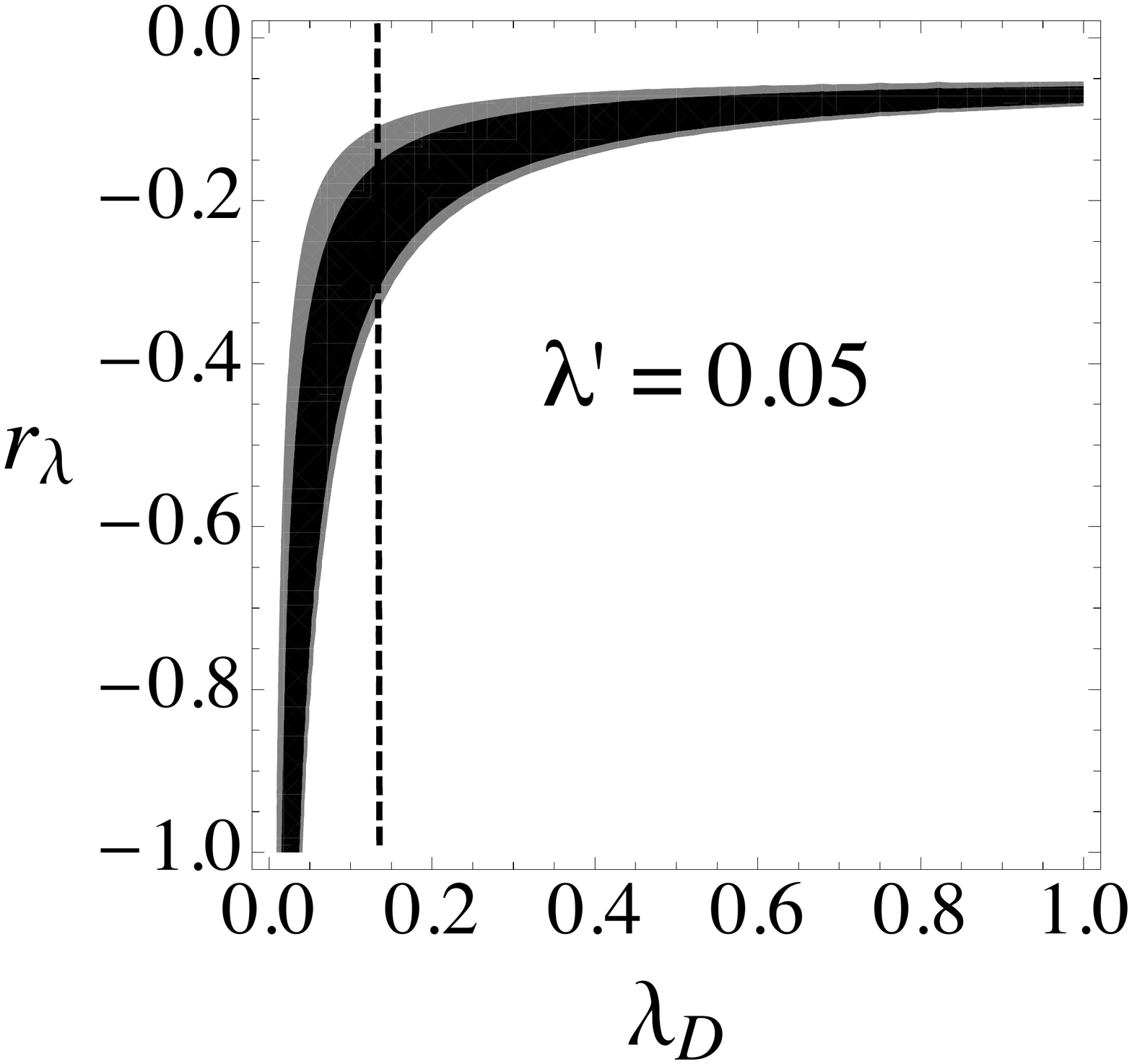}
\includegraphics[width=5cm,height=6.5cm,angle=0]{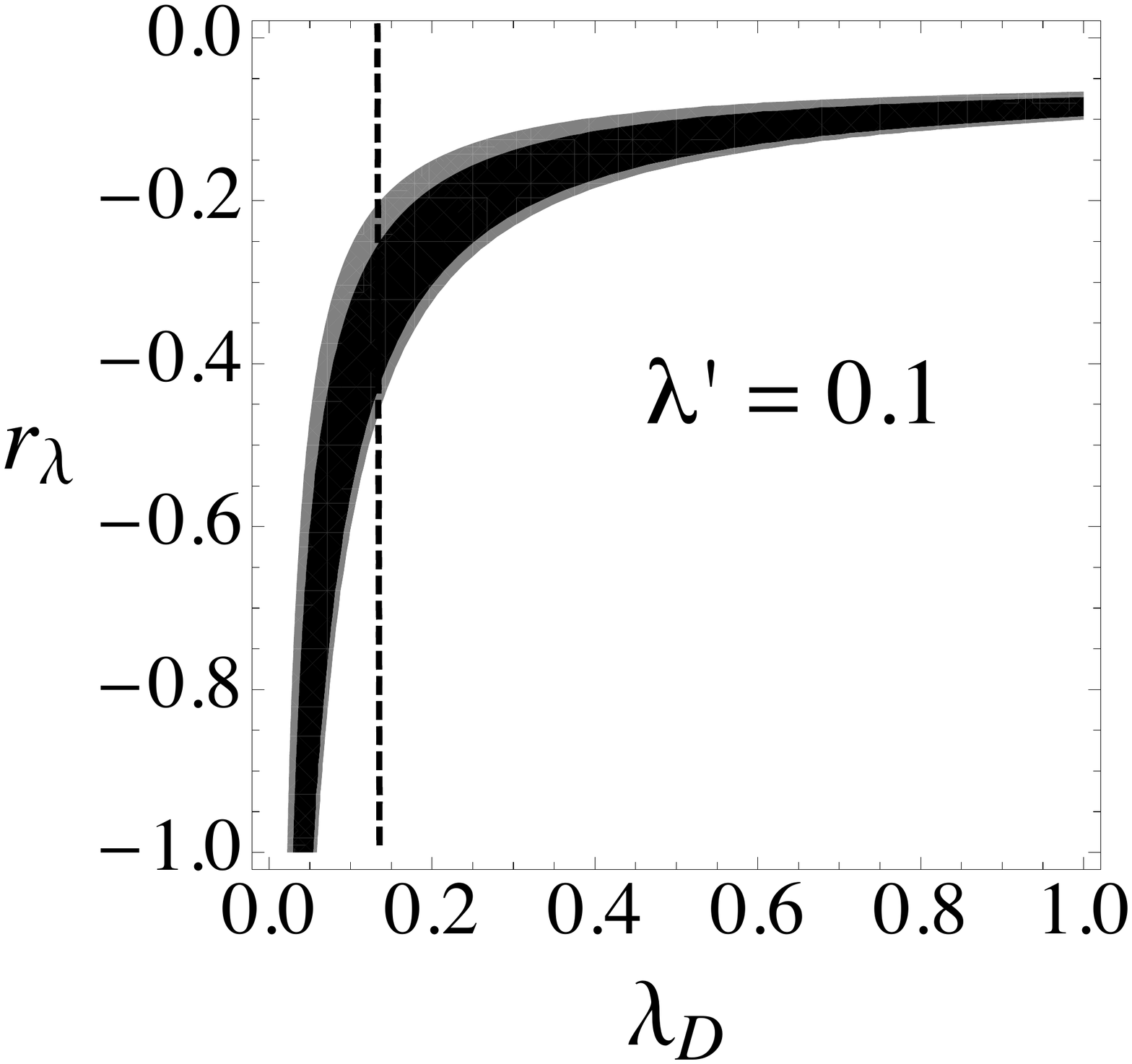}
\includegraphics[width=5cm,height=6.5cm,angle=0]{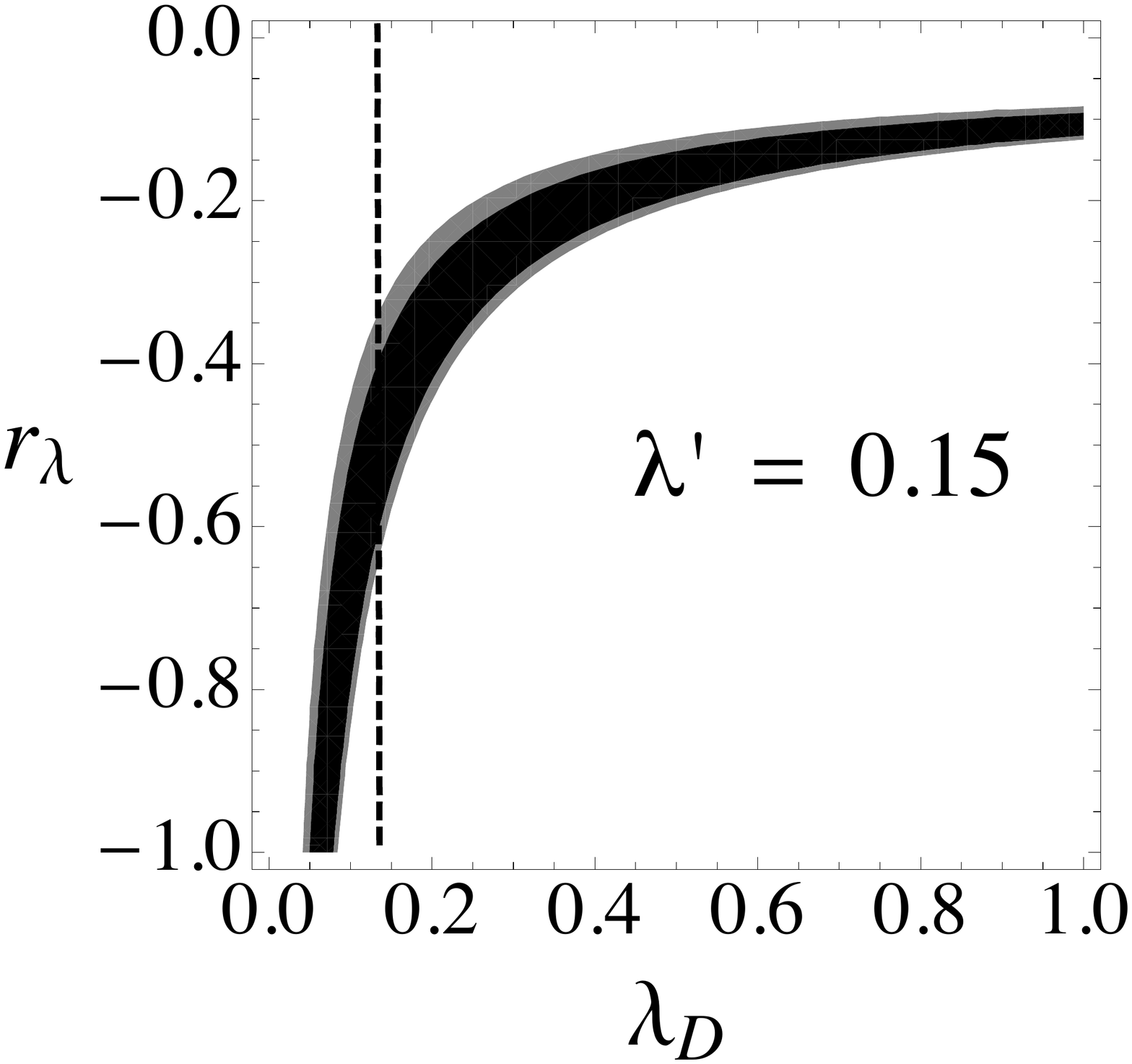}\vspace{-1cm}
\caption{Allowed regions in the $(\lambda _{D}, r_{\lambda })$ plane, compatible with the diphoton Higgs decay limits at ATLAS (black region) and CMS (gray region), for three values of $\lambda '$ . The other parameters are fixed to be: $M_{H^{\pm }}=300$ GeV, $M_{\sigma _{0}}=60$ GeV and $T_{\beta }=10$.  The dashed line shows the mass constraint from (\ref{higgs-mass-2})}
\label{fig5}
\end{figure}

\begin{figure}[tbh]
\centering
\includegraphics[width=8.5cm,height=6.5cm,angle=0]{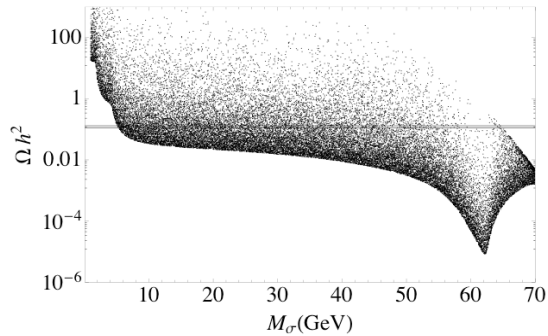}
\caption{Relic density as function of $M_{\sigma}$ with $0.02\leq\lambda '\leq 0.08$ and $T_{\beta }=10$, $M_{H_{0 }}=300$ GeV, $r_{\lambda }=-0.3$ and $\lambda _{D}=0.13$.}
\label{fig6}
\end{figure}

If we assume that $\sigma _0$ is the single DM component of the Universe, we can calculate the DM relic density from all dispersions of $\sigma _0$ into the ordinary SM matter, as described in detail in reference \cite{U1DM}. Taking into account that the current limit for DM relic density is $\Omega h^2=0.1199\pm 0.0027$ \cite{planck}, we obtain in Fig. \ref{fig6} limits on the mass of $\sigma _0$ for the parameters used in Fig. \ref{fig4}, with  $T_{\beta }=10$, $r_{\lambda}=-0.3$, $\lambda_D=0.13$, and $M_{H^{\pm}}=M_{H_0}=300$ GeV. According to Fig. \ref{fig4}, we use for $\lambda '$ random values between $0.02$ and $0.08$. First, we see that for very small values of $M_{\sigma}$ (below $5$ GeV), large DM relic density is obtained, with values above the experimental limit described by the horizontal band. Above this limit, we find points into the experimental band. However, when the mass increases, there arises a resonance around $M_{\sigma}=62$ GeV due to the pole $(2M_{\sigma})^2-M_{h_{0}}^2$, which produces an excess of DM annihilation through the process $\sigma,\sigma \rightarrow h_0 \rightarrow SM,SM$ mediated by the SM-like Higgs boson. At the resonance, we only see few points consistent with the observed relic density, which corresponds to the smallest allowed value of the DM-Higgs coupling $\lambda '\sim 0.02$. For $M_{\sigma}>64$ GeV, we do not find any allowed point, and the invisible Higgs decay into DM is forbidden.

 \subsection{Direct detection}
     
\begin{figure}[tbh]
\centering
\includegraphics[width=9cm,height=7cm,angle=0]{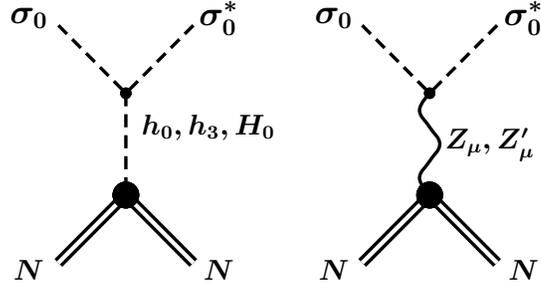}
\caption{DM-nuclei scattering mediated by t-channel exchange of Higgs bosons and neutral gauge bosons.}
\label{fig7}
\end{figure}

\begin{figure}[tbh]
\centering
\includegraphics[width=8cm,height=7cm,angle=0]{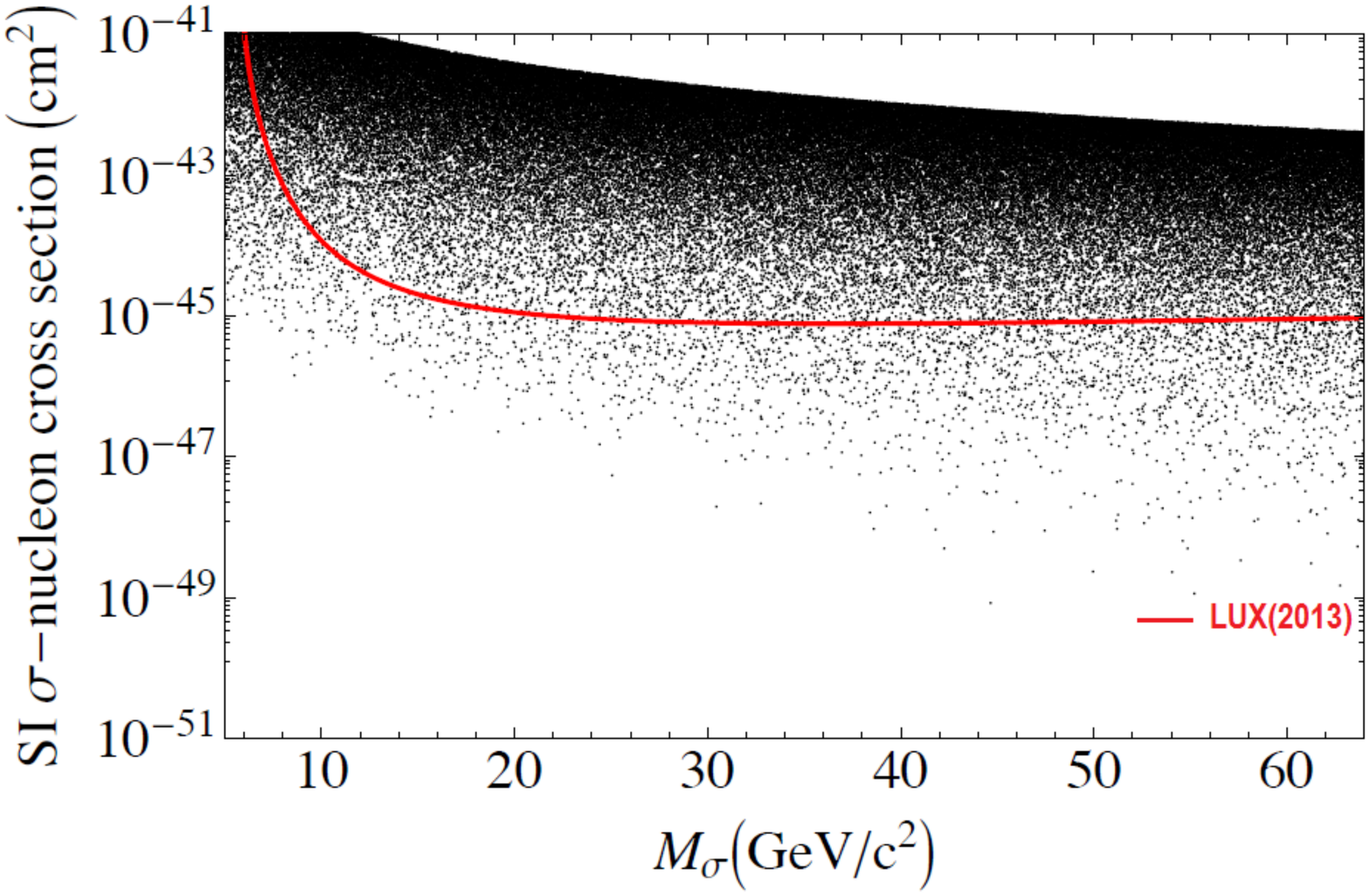}
\includegraphics[width=8cm,height=6.5cm,angle=0]{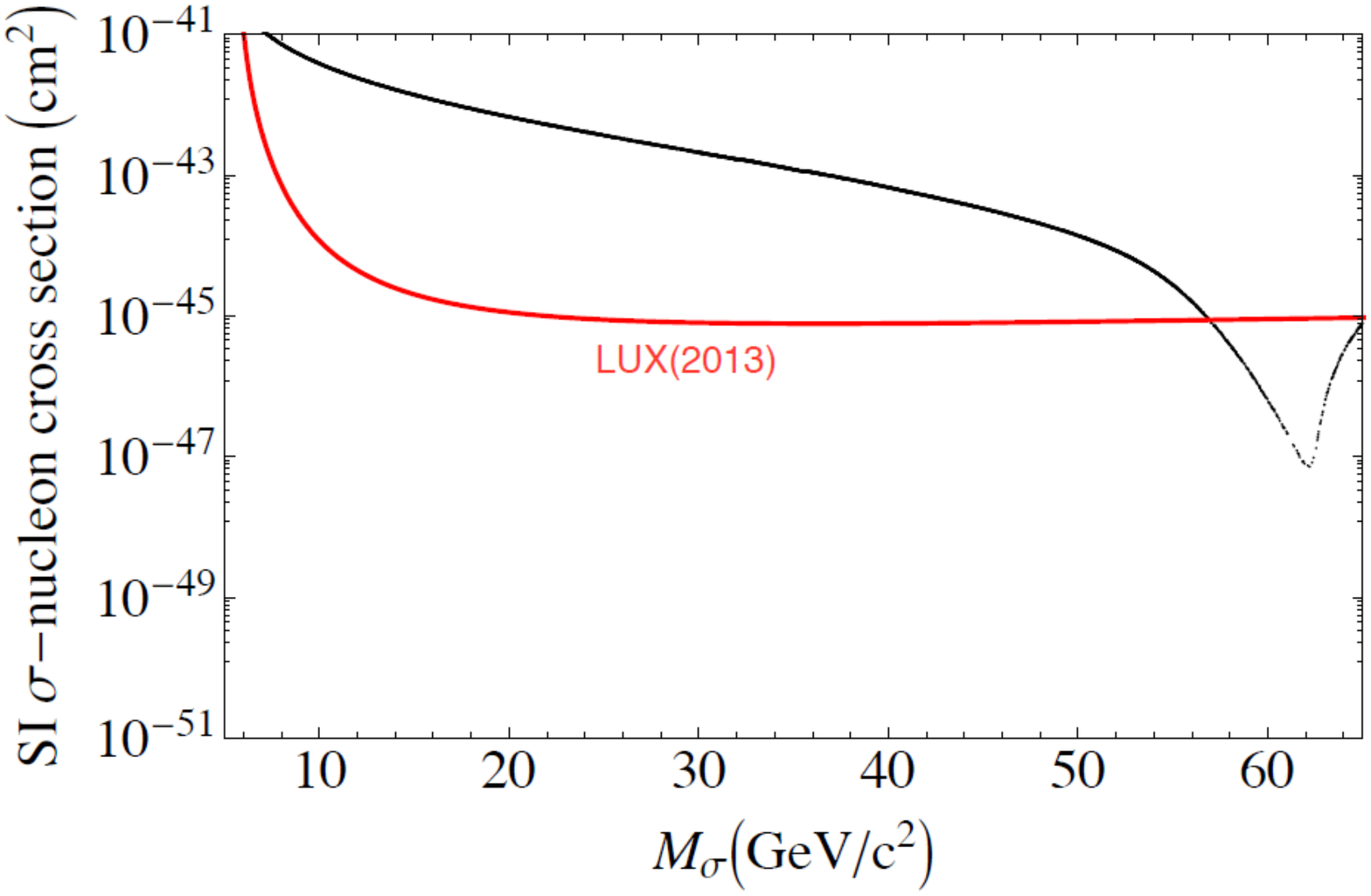}
\caption{Spin-independent cross section for DM-proton scattering. On the left, the scan satisfies constraints only from diphoton Higgs decay limits, and on the right, the points satisfy in addition the relic density bound. Here $T_{\beta }=10$, $M_{Z'}=10000$ GeV, and the couplings of the heavy Higgs bosons are set to be zero}
\label{fig8}
\end{figure}

For DM masses above 6 GeV, the LUX collaboration \cite{lux} has obtained strong limits on the spin-independent DM-nucleus elastic cross section, which impose additional constraints to the parameters of the model. In our case, the DM scatter off nuclei through t-channel exchange of Higgs and neutral gauge bosons, as shown in Fig. \ref{fig7}. 
Since the couplings of scalars to ordinary matter are the same as in a two-Higgs doublet model \cite{U1,U1DM}, and in order to avoid excess of flavor changing neutral processes, we use the parameters of a THDM type II in the micrOMEGAs to describe the couplings of the Higgs bosons with the quark structure of the nucleons, while the couplings of the gauge bosons and quarks are described in ref. \cite{U1DM}.  First, to study the effects of the constraints from the diphoton Higgs decay and relic density, we take a large value for the mass of the $Z'$ boson, $M_{Z'}=10000$ GeV, and set the couplings of the others Higgs bosons to be zero. We obtain in Fig. \ref{fig8} on the left the cross section as a function of the DM mass, where we select random points that only satisfy the diphoton contraints. The experimental limit from LUX is also shown, where only those points below the line satisfy the LUX limit. The plot on the right show points that satisfy in addition the relic density constraint. In this case, for DM masses below 57
GeV, the data does not fit the LUX limit. However, to reproduce the observed relic density around the region with $M_{\sigma _0}=M_{h_0}/2\approx 62$ GeV, we must keep small values of the coupling $\lambda '$, near the lower limit of 0.02, to compensate the Higgs resonance in the cross section for DM annihilation. Thus, we obtain a large suppression of the DM scattering cross-section around this region, as shown in the figure, obtaining allowed points for $M_{\sigma _{0}}\geq 57
$ GeV. 

\begin{figure}[tbh]
\centering
\includegraphics[width=8.5cm,height=6.5cm,angle=0]{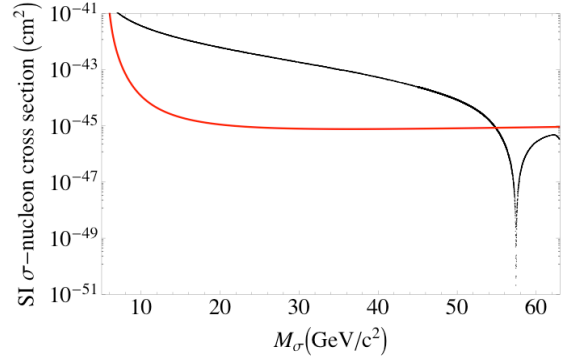}
\caption{Spin-independent cross section for DM-proton scattering that satisfies constraints from diphoton Higgs decay limits and relic density bounds. Here $T_{\beta }=10$, $M_{Z'}=3000$ GeV, and we consider couplings of the DM with all the Higgs sector.}
\label{fig9}
\end{figure}

On the other hand, we obtain a similar region if we consider lower masses for the $Z'$ gauge boson, for example, $3000$ GeV, and take into account interactions with the others Higgs bosons, where a large suppresion near the Higgs resonance arises, as shown in Fig. \ref{fig9}. However, due to the quantum interference among the Higgs bosons and the $Z'$ channels, the maximum suppresion shift to $M_{\sigma _0}\approx 57$ GeV. In this case, the allowed region is above $M_{\sigma _{0}}\geq 55
$ GeV.  

\section{The trilineal self-coupling\label{sec:trilineal}}

The trilineal Higgs self-coupling will be an important parameter to be measured at LHC in order to prove the consistence between the observed and the SM Higgs boson \cite{trilineal}. In our model, there arise scenarios where the trilineal coupling  exhibits new physics deviation from the SM prediction. As shown in Eq. (\ref{trilineal-coup}), the trilineal coupling $g_{3h}$ is function of the same parameters that contibute to the Higgs diphoton decay. Thus, the diphoton constraints can be used in order to estimate the deviation of the Higgs self-coupling from the SM prediction. Using the same parametrization as in (\ref{theoretical-constraint-reduce}), we obtain for the trilineal self-coupling in (\ref{trilineal-coup}) that:

\begin{eqnarray}
g_{3h}&=&6\upsilon \lambda _D\left[C_{\beta }C_{\alpha }^{3}+S_{\beta }S_{\alpha }^{3}+r_{\lambda }\left(C_{\beta }S_{\alpha }+S_{\beta }C_{\alpha }\right)C_{\alpha }S_{\alpha }\right].
\label{trilineal-coup-2}
\end{eqnarray}
By expanding to dominant order the relation between the rotation angles from Eq. (\ref{angle-relations-x}), we obtain that:

\begin{eqnarray}
g_{3h}&\approx&6\upsilon \lambda _D\left[1-\frac{1}{2}S_{2\beta }^2(1-r_{\lambda })+\frac{\upsilon ^2}{2M_{H^{\pm }}^2}C_{2\beta }^2S_{2\beta }^2\lambda _D(1-r_{\lambda })(2r_{\lambda }-3)\right]\nonumber \\
&=&\frac{3M_{h_0}^2}{\upsilon }\left[1+\frac{\upsilon ^4}{M_{h_0}^2M_{H^{\pm }}^2}C_{2\beta }^2S_{2\beta }^2\lambda _D^2(1-r_{\lambda })(2r_{\lambda }-3)\right]
\label{trilineal-coup-3}
\end{eqnarray}
where we applied the relation for $M_{h_0}$ from (\ref{higgs-mass-2}). In the SM, the trilineal self-coupling is:

\begin{eqnarray}
g_{3h}^{SM}=\frac{3M_{h_0}^2}{\upsilon }.
\end{eqnarray}
Thus, by defining the fractional deviation as $\delta g=(g_{3h}-g_{3h}^{SM})/g_{3h}^{SM}$, we obtain from (\ref{trilineal-coup-3}) that:

\begin{eqnarray}
\delta g=\frac{\upsilon ^4}{M_{h_0}^2M_{H^{\pm }}^2}C_{2\beta }^2S_{2\beta }^2\lambda _D^2(1-r_{\lambda })(2r_{\lambda }-3).
\label{trilineal-deviation}
\end{eqnarray}
Since $r_{\lambda }$ must take values below 1, the factor $(1-r_{\lambda })(2r_{\lambda }-3)$ is always negative. Thus, the deviation is negative, which implies that $g_{3h}\leq g_{3h}^{SM}$. On the other hand, we identify three scenarios for the decoupling limit of the trilineal self-coupling, where $g_{3h}= g_{3h}^{SM}$ : 

\begin{enumerate}
\item $2\beta =0$, $\pi /2$ and $\pi $. These angles lead us to $T_{\beta }=0, 1$ and going to infinity.

\item $r_{\lambda }=1$. In this decoupling limit, all the Higgs doublet couplings are the same ($\lambda _D=\lambda _5=\lambda '_5$). 

\item $M_{H^{\pm }}^2 \gg \upsilon ^2$, which corresponds to the traditional large-mass decoupling limit.
\end{enumerate}

We see that Eq. (\ref{trilineal-deviation}) exhibits another decoupling limit when $r_{\lambda }=3/2$. However, due to unitarity and stability, the parameter $r_{\lambda}$ can not be larger than $1$.

\begin{figure}[tbh]
\centering
\includegraphics[width=6.5cm,height=8.5cm,angle=0]{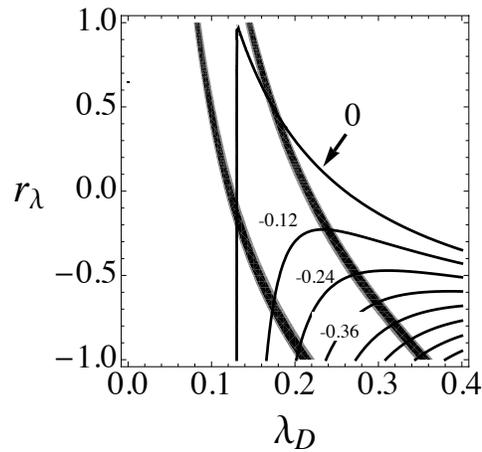}\vspace{-1cm}
\caption{Contour plots of $\delta g$ with -0.12 line spacing starting from $0$. The bands corresponds to the Higgs dipoton decay constraints.}
\label{fig10}
\end{figure}

On the other hand, the deviation in (\ref{trilineal-deviation}) is function of four unknown parameters: $(T_{\beta }, \lambda _D, r_{\lambda }, M_{H^{\pm }})$. However, due to the analytical expression for the Higgs mass in (\ref{higgs-mass-2}), we can reduce the number of free parameters into three. For example, if we solve for $T_{\beta}$ as function of $(\lambda _D, r_{\lambda})$ and by fixing $M_{H^{\pm }}=300$ GeV, we can plot contour plots of $\delta g$ in the $(\lambda _D, r_{\lambda})$ plane. Fig. \ref{fig10} shows equally spaced contour plots. The line spacing is set in -0.12, starting from $\delta g =0$. We also show the allowed band region from the Higgs diphoton decay data. The largest deviation compatible with the diphoton constraint is $\delta g =-72\%$, around $\lambda _D=0.35$ and $r_{\lambda}=-1$. Above $\delta g =-24\%$, the contour plots exhibit two solutions: first around $(\lambda _D,r_{\lambda})=(0.27,-0.5)$, and later near $(\lambda _D,r_{\lambda})=(0.2,-1)$. There are also solutions compatible with null deviation ($\delta g =0$) around $(\lambda _D,r_{\lambda})=(0.17, 0.5 )$ and $(\lambda _D,r_{\lambda})=(0.13,-0.13)$.          

\section{Conclusion\label{sec:conclusions}}

Constraints on the scalar potential couplings of a nonuniversal $U(1)'$ extension of the SM were obtained using unitarity and stability of the Higgs potential. Using recent data from CERN-LHC collider, we obtain allowed points of the scalar parameters compatible with the signal strenght of diphoton Higgs decay. We conclude that:

\begin{enumerate}
\item By combining the unitarity and stability conditions, we obtain individual bounds and order relations between coupling constants. In particular, the scalar interactions of the observed $125$ GeV Higgs boson depends on the six parameters $\lambda _{1, 2, 5}$, $\lambda '_{5, 6, 7}$. The theoretical constraints impose positive bounds on $\lambda _{1,2}$, while $\lambda _{5}$, $\lambda '_{5}$ and $\lambda _{1,2}$ holds specific order relations. On the other hand, the couplings associated with the scalar DM candidate, $\lambda '_{6,7}$, are only bounded from below.

\item The observed diphoton Higgs decay at LHC imposes phenomenological constraints on the above couplings as well as on other scalar parameters. Since the signal strenght depends on the total decay of the Higgs boson there arises two possible scenarios. In the first one, decays into DM is forbidden for masses above $M_{h_{0}}/2 \approx 63$ GeV. By assuming that the lightest Higgs boson corresponds to the observed one, we obtain very stringent bounds on the free parameters. In particular, small and negative values of the ratio $r_{\lambda}$ is favoured, which implies that the mixing couplings between Higgs doublets are suppresed as compared with the diagonal couplings. Since the diagonal coupling $\lambda _D$ must be positive because of the stability of the Higgs potential, negative mixing couplings is favoured by the experimental data from diphoton decay. Also, large values on $T_{\beta }$ above 5 exhibits larger allowed intervals. Finally, we observe that the allowed points is not very sensitive to the heavy scale $M_{H^{\pm }}$.

\item In the scenario where the mass of the DM candidate is below the kinematical treshold of $63$ GeV, the signal strenght become sensitive to the couplings $\lambda '_{6,7}=\lambda '$ and the mass $M_{\sigma _{0}}$. We found allowed intervals consistent with LHC data on the diphoton decay in an scenary where invisible decay into a light DM candidate is possible.

\item  Using the above diphoton constraints, we evaluate the allowed masses for DM compatible with the observable relic density of the Universe. For the range $0.02 \leq \lambda ' \leq 0.08$ from Fig. \ref{fig4}, we found allowed masses in the range $5$ GeV $\leq M_{\sigma} \leq 62$ GeV for light scalar DM, with only a few allowed points around the resonance at $62$ GeV.

\item Direct detection from the elastic DM-nuclei dispersion mediated by the t-channel Higgs exchange was evaluated and compared with the experimental limits from LUX collaboration. Using parameters compatible with the diphoton Higgs decay and the observed DM relic density, we obtain the cross section as a function of the DM mass. The LUX limit discard light DM below 57 GeV for $M_{Z'}=10000$ GeV, and below 55 GeV for $M_{Z'}=3000$ GeV. 

\item We evaluate deviations of the trilineal Higgs self-coupling from the SM prediction. By combining the unitarity, stability and the Higgs diphoton decay constraints, we found allow trilineal deviations in the range $0 \leq \delta g \lesssim -72\%$. 
    
\end{enumerate}

\section*{Acknowledgment}

This work is supported by El Patrimonio Aut\'onomo Fondo Nacional de Financiamiento para la Ciencia, la Tecnolog\'{i}a y la Innovaci\'on Fransisco Jos\'e de Caldas programme of COLCIENCIAS in Colombia.

\newpage

\begin{table}[]
\begin{center}
\caption{\small Ordinary SM particle content, with $i=$1,2,3} \vspace{-0.5cm}
\label{tab:SM-espectro}
\begin{equation*}
\begin{tabular}{c c c c}
\hline\hline
$Spectrum$ & $G_{sm}$ & $U(1)_{X }$ & $Feature$\\ \hline \\
$\
\begin{tabular}{c}
$q^i_{L}=\left(
\begin{array}{c}
U^i \\
D^i
\end{array}%
\right) _{L}$ 
\end{tabular}%
\ $ &
$(3,2,1/3)$
&
$\
\begin{tabular}{c}
$1/3$ for $i=3$ \\
$0$ for $i=1,2$%
\end{tabular}
\ $
&
chiral
\\ \\
$U_R^i$
&
$(3^*,1,4/3)$
&
\begin{tabular}{c}
$2/3$ 
\end{tabular}
&
chiral
\\ \\
$D_R^i$
&
$(3^*,1,-2/3)$
&
\begin{tabular}{c}
$-1/3$ 
\end{tabular}
&
chiral
\\ \\
$\
\begin{tabular}{c}
$\ell ^i_{L}=\left(
\begin{array}{c}
\nu ^i \\
e^i
\end{array}%
\right) _{L}$ \\
\end{tabular}
\ $
&
$(1,2,-1)$
&
$-1/3$
&
chiral
\\  \\
$e_R^i$
&
$(1,1,-2)$
&
\begin{tabular}{c}
$-1$ 
\end{tabular}
&
chiral
\\  \\
$\
\begin{tabular}{c}
$\phi _{1}=\left(
\begin{array}{c}
\omega _1^{+} \\
\frac{1}{\sqrt{2}}(\upsilon _1+h_1+iz_1)
\end{array}%
\right) $
\end{tabular}
\ $
&
$(1,2,1)$
&
$2/3$
&
Scalar Doublet
\\  \\
$\
\begin{tabular}{c}
$W _{\mu }=\left(
\begin{array}{cc}
W _{\mu }^{3}&\sqrt{2}W_{\mu }^+ \\
\sqrt{2}W_{\mu }^-& -W _{\mu }^{3}
\end{array}%
\right) $
\end{tabular}
\ $
&
$(1, 2 \times 2^*, 0)$
&
$0$
&
Vector
\\  \\
$\
B_{\mu }
\ $
&
$(1, 1, 0)$
&
$0$
&
Vector
\end{tabular}%
\end{equation*}%
\end{center}
\end{table}

\begin{table}[tbp]
\begin{center}
\caption{\small Exotic non-SM particle content, with $n=$1,2}\vspace{-0.5cm}
\label{tab:exotic-espectro}
\begin{equation*}
\begin{tabular}{c c c c}
\hline\hline
$Spectrum$ & $G_{sm}$ & $U(1)_{X }$ & $Feature$\\ \hline \\
$\
\begin{tabular}{c}
$T_L$
\end{tabular}%
\ $ &
$(3,1,4/3)$
&
$1/3$
&
quasi-chiral
\\ \\
\begin{tabular}{c}
$T_R$
\end{tabular}
&
$(3^*,1,4/3)$
&
$2/3$
&
quasi-chiral
\\ \\
\begin{tabular}{c}
$J_L^n$
\end{tabular}
&
$(3,1,-2/3)$
&
$0$
&
quasi-chiral
\\ \\ 
$\
\begin{tabular}{c}
$J_R^n$
\end{tabular}
\ $
&
$(3^*,1,-2/3)$
&
$-1/3$
&
quasi-chiral
\\ \\
\begin{tabular}{c}
$(\nu _R^i)^c$
\end{tabular}
&
$(1,1,0)$
&
$-1/3$
&
Majorana
\\ \\ 
\begin{tabular}{c}
$N _R^i$
\end{tabular}
&
$(1,1,0)$
&
$0$
&
Majorana
\\ \\
$\
\begin{tabular}{c}
$\phi _{2}=\left(
\begin{array}{c}
\omega _2^{+} \\
\frac{1}{\sqrt{2}}(\upsilon _2+h_2+iz_2)
\end{array}%
\right) $
\end{tabular}
\ $
&
$(1,2,1)$
&
$1/3$
&
Scalar doublet
\\ \\
\begin{tabular}{c}
$\chi _0 = \frac{1}{\sqrt{2}}(\upsilon _{\chi}+h_{3 }+iz_{3})$
\end{tabular}
&
$(1,1,0)$
&
$-1/3$
&
Scalar singlet
\\ \\
\begin{tabular}{c}
$\sigma _0 = \frac{1}{\sqrt{2}}(\upsilon _{\sigma}+h_4+iz_4)$
\end{tabular}
&
$(1,1,0)$
&
$-1/3$
&
Scalar singlet
\\  \\ 
$Z'_{\mu}$ 
&
$(1,1,0)$
&
$0$
&
Vector
\end{tabular}%
\end{equation*}%
\end{center}
\end{table}

\end{document}